\documentclass[aps,prb,floatfix,amssymb,amsfonts,amsmath,reprint,superscriptaddress]{revtex4-2}
\usepackage[utf8]{inputenc}
\usepackage[T1]{fontenc}
\usepackage{amsfonts}
\usepackage{graphicx}
\usepackage{amssymb}
\usepackage{amsmath}
\usepackage{dcolumn}
\usepackage{bm}
\usepackage{textcomp}
\usepackage{xcolor}
\usepackage{wasysym}
\usepackage{flushend}

\def\kHg{ET-HgCl }
\def\ET{BEDT-TTF }
\def\cm{cm$^{-1}$~}

\begin{document}

\title{Raman signatures of the strong  intra- and inter-molecular charge oscillations\\ in bis(ethylenedithio)-tetrathiafulvalene (BEDT-TTF) $\kappa$-phase salts}
\author{A. Girlando}
\affiliation{Molecular Materials Group (MoMaG), 43124 Parma, Italy}

	\begin{abstract}
First principle calculations of the Raman intensities of a
(BEDT-TTF)$_2^+$ centrosymmetric dimer lead to a full reconsideration
of the assignment of the three C=C stretching phonons in
$\kappa$-phase BEDT-TTF salts. In contrast with previous
interpretation, it is found that the presence of 
the out-of-phase coupling of the antisymmetric external C=C stretching mode
has also to be taken into account. This mode, infrared-active for a single BEDT-TTF
molecule, implies a strong intra-molecular charge oscillation along BEDT-TTF long 
molecular axis. In the consequent reassignment of the C=C spectral
region, a very broad band appearing
in the cross polarized Raman spectra is interpreted as due to \textit{inter}-dimer
electron-molecular vibration (e-mv)
coupling, to be contrasted with the well known intra-dimer
e-mv coupling that induces very strong infrared absorptions.
Analysis of the data in term of a properly developed e-mv scheme
yields an evaluation of the Hubbard parameters relevant
to the so-called effective dimer model often used to interpret
the physical properties of $\kappa$-BEDT-TTF salts. Finally
the Raman spectra of an asymmetric 
(BEDT-TTF)$^{+0.6}$(BEDT-TTF)$^{+0.4}$ dimer have been calculated in order
to interpret the Raman spectra of the intriguing charge-ordered
(CO), ferroelectric phase of $\kappa$-(BEDT-TTF)$_2$ Hg(SCN)$_2$Cl.
It turns out that in the asymmetric dimer
the strong intra-dimer the e-mv induced infrared C=C stretching absorption should
appear with huge intensity also in Raman.
The absence of such band in the spectra of $\kappa$-(BEDT-TTF)$_2$ Hg(SCN)$_2$Cl
ferroelectric phase gives clear alternative indications about the possible CO pattern
in such a phase.
\end{abstract}
	
\maketitle
\section{Introduction}
The salts of bis(ethylenedithio)-tetrathiafulvalene (BEDT-TTF, ET for short) molecule and its variants have played and play a central role among molecular quantum materials. Besides superconductivity, many other interesting quantum phenomena are occurring in these salts, such as charge fluctuations, inter-molecular charge-ordering and ferroelectricity, spin liquids state and so forth \cite{Powell2011,Ardavan2012,Wosnitza2012,Dressel2020,Riedl2022}. Most salts have stoichiometry (\ET)$_2^+$X$^-$ (X is a closed-shell anion), namely an electron every two \ET, and the \ET molecules are packed 
roughly upright, forming layers separated by sheets of anions or anion polymeric networks. The structure and physical properties are then quasi two-dimensional (2D), and the interesting physics lye in the
ET layers, with anions sheets just considered to serve as spacer and charge reservoir - although they may have a role in stabilizing the diverse electronic phases of the donor layer.  The ET molecules can arrange in several different ways within the layers, even with the same counterions, yielding polymorphism with associated variety of physical properties \cite{Mori1984,Mori1998,Mori1999}, the different molecular packings being labeled by different Greek letters. The sharing of one electron between two molecular units favor the arrangement of ET in pairs, or dimers, within the layers. The degree of dimerization changes
from almost nil in the case of the $\theta$-phase, to the chessboard arrangements of dimers of the $\kappa$-phase. 

\begin{figure}
	\centering
	\includegraphics[width=0.95\linewidth]{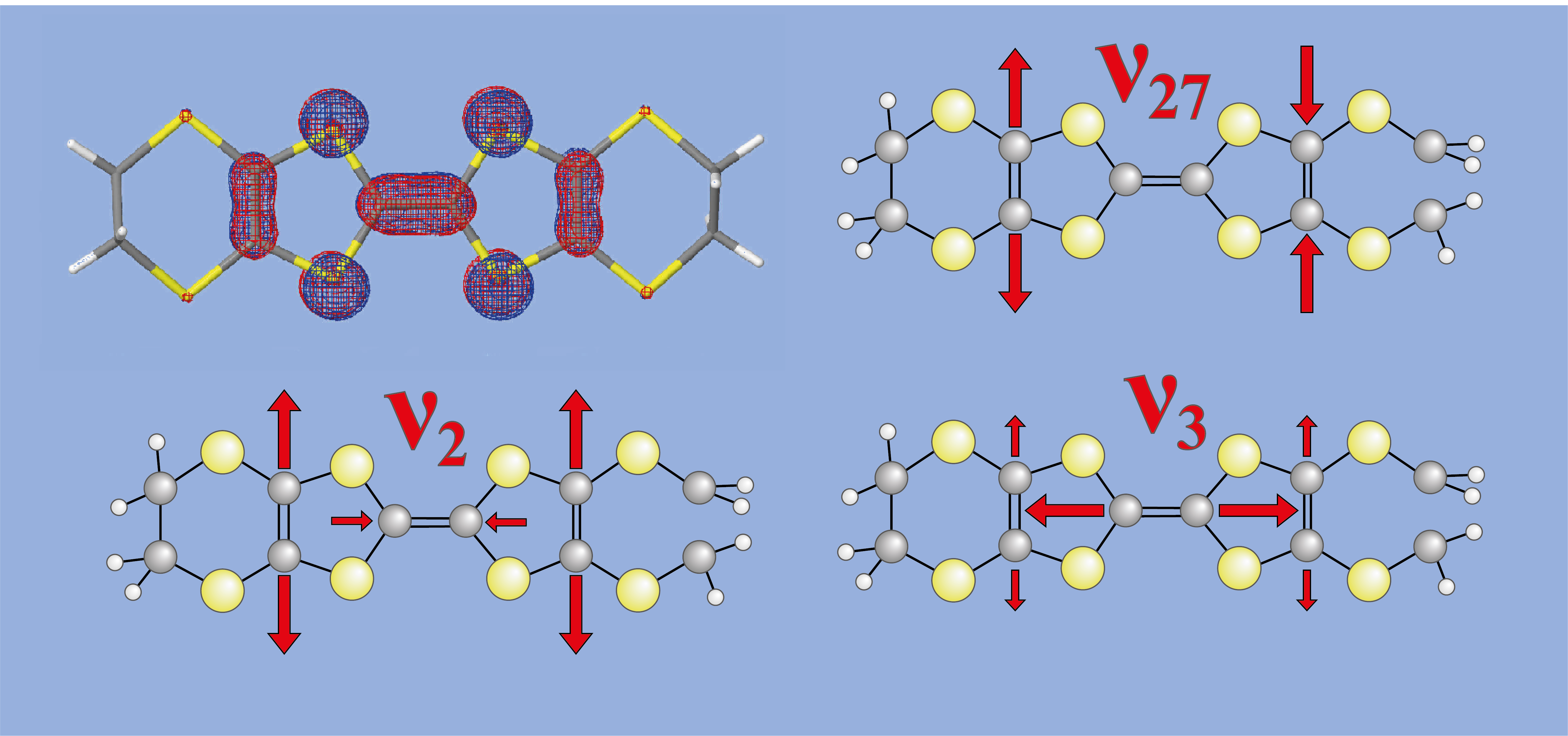}
	\caption{BEDT-TTF HOMO orbital (upper left corner) and the three C=C stretching phonons, $b_{1u}~ \nu_{27}$, $a_g ~\nu_2$ and $\nu_3$.}
	\label{fig:ET_C=Cvibs}
\end{figure}

ET salts are characterized by strongly correlated mobile electrons, and the interplay of
competing but comparable interactions yield to several types of instabilities. Among the
many different techniques used to probe the ET salts physical properties and investigate the
resulting complex ``phase diagram'', a prominent place is played by vibrational spectroscopy, due to the
strong coupling between electrons and phonons. In the case of ET, the most important phonons from this point of view are the three C=C stretchings, just because the HOMO orbital is mainly localized
on the corresponding bonds (Fig. 1). Therefore the removal of one electron weakens the bond,
with large and practically linear downshift of the associated phonon frequency ($\gtrsim$ 100 \cm), yielding estimate of the average charge residing on the molecule \cite{Yamamoto2005,Girlando2011}.
Moreover, the two totally symmetric modes modulate the HOMO, and are then directly coupled to
the electron (e-mv coupling), giving rise to very strong infrared (IR) bands. The comparison between Raman and infrared (IR) spectra thus yields an estimate of the e-mv coupling constants \cite{Rice1979,Painelli1986}.
In addition, analysis of the band shape of the C=C stretching vibrations, notably the IR
active $b_{1u}$ $\nu_{27}$ (cf. Fig. \ref{fig:ET_C=Cvibs}), has yielded an estimate of the charge fluctuation velocity between ET molecules \cite{Girlando2012,Girlando2014}.

While both IR and Raman spectroscopy generally provide useful and complementary information, in case of ET salts most attention, both theoretical and experimental, has been devoted to IR. There are several reasons for that, the main one being that measuring and calculating Raman intensities is complicated. There is also the problem of resonance or pre-resonace effects \cite{Maksimuk2001,Yamamoto2021}, with associated radiation absorption, fluorescence, etc. First principles Density Functional (DFT) calculations were then limited to the IR intensities \cite{Girlando2011}, and until recently just one paper tried to assign the Raman active C=C stretching of $\kappa$-phase ET salts through polarization measurements, isotopic substitution
and the use of different excitation lines and temperatures \cite{Maksimuk2001}. Despite this outstanding effort, some interpretation remained uncertain (cf. Table III of the latter paper).    

Recent high quality polarized Raman spectra of $\kappa$-(BEDT-TTF)$_2$ Hg(SCN)$_2$Cl (\kHg for short) \cite{Hassan2020}, a salt undergoing a intriguing re-entrant charge order (CO) phase transition \cite{Drichko2014,Loehle2016,Hassan2020},  have shown that the C=C stretching spectral region is more complex than generally thought.
Considering the strong dimerization of $\kappa$-phase salts, and of this one in particular \cite{Gati2018}, I
decided to perform DFT calculations of (BEDT-TTF)$_2^+$ dimers along the lines sketched in Ref. \cite{Girlando2011}, this time including calculation of the (non-resonant) Raman intensities of dimers keeping the same orientation as in the crystal, in order to reproduce the polarized measurement. Qualitative but quite satisfactory agreement with experiment, also considering the above
sketched difficulties of assessing Raman intensities, prompts for a full re-consideration of
the assignment of the Raman active C=C stretching phonons in $\kappa$-phase salts.
Furthermore, the presence of a very broad band in the cross-polarized spectra is explained in terms
of an effect of e-mv coupling involving inter-dimer interaction, allowing an evaluation of
relevant Hubbard parameters. Finally, DFT calculation of a charge-ordered dimer
gives important clues about the possible CO pattern in $\kappa$-(BEDT-TTF)$_2$ Hg(SCN)$_2$Cl ferroelectric phase \cite{Gati2018}, and in others
$\kappa-$phase salts displaying similarly originated phenomena \cite{Riedl2022}.

\section{Methods }

\subsection{Spectral predictions}

Table I reports the spectral predictions for 
the crystal phonons, using the correlation diagram that starts from the isolated molecule
\cite{Turrell1972}. BEDT-TTF is not planar ($D_2$ symmetry \cite{Girlando2011}), but since here we are concerned with the three C=C stretchings, which constitute the planar part of the molecule, I shall follow the usual convention of 
classifying these phonons in terms of $D_{2h}$ symmetry: for the isolated molecule,
two stretching  modes are Raman active ($a_g~\nu_3$ and $\nu_2$), and one is IR active
($b_{1u}~\nu_{27}$). A cartoon of the three vibrations is reported in Fig. \ref{fig:ET_C=Cvibs}.
In the crystal, the site symmetry of each ET is $C_1$, and in principle all the modes are
both Raman and IR active. But the deviation from the isolated molecule symmetry is very small,
so one expects that isolated molecule selection rule still apply.
The two ET molecules in the dimer are connected by an inversion center, and each ET vibration is coupled
in-phase and out-of-phase: For the C=C stretchings three IR active ($\mathcal{A}_u$) and three
Raman active ($\mathcal{A}_g$) modes are expected. The three IR active modes have been identified: one is charge sensitive mode corresponding to the in-phase combination of the $b_{1u}~\nu_{27}$, and the
other two are the e-mv coupled out-of-phase combination of $a_g~\nu_2$ and $\nu_3$ \cite{Girlando2011}.
On the other hand, in Raman the C=C stretching region is dominated by two strong bands, up to now
identified as the in-phase combination of the $a_g~\nu_2$ and $\nu_3$, whereas the assignment of the
$b_{1u}~\nu_{27}$ remains uncertain \cite{Maksimuk2001}.

\begin{figure*}
	\centering
	\label{correlationtable}
	\includegraphics[width=0.8\linewidth]{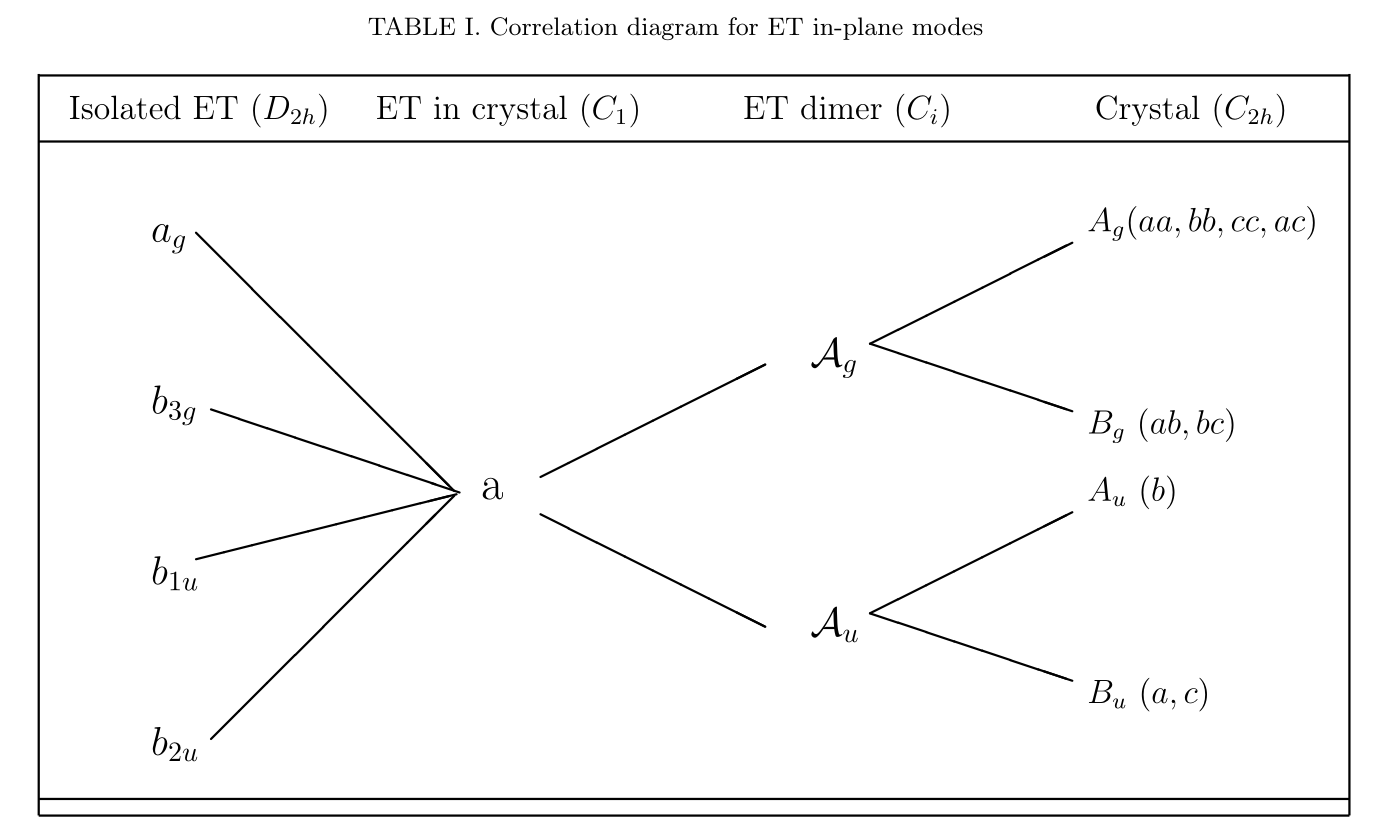}
\end{figure*}

\begin{figure}
	\centering
	\includegraphics[width=0.9\linewidth]{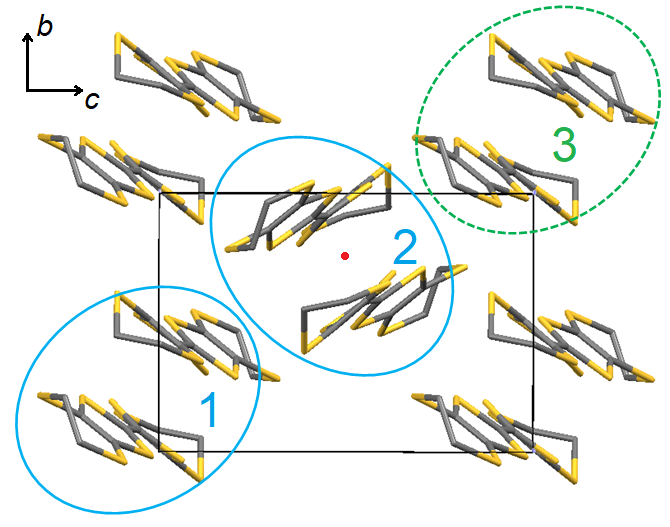}
	\caption{Organic layer ($bc$ plane) of \kHg at 50 K. The continuous light blue ellipses 
		indicate the two dimers in the unit cell, the green dashed line referring to the third
		dimer in the trimer cluster used for modeling the e-mv interaction. The red dot 
	indicate the inversion center for both the dimer 2 and for the supermolecule trimer 123.}
	\label{fig:kappaetlayer}
\end{figure}

The spectral prediction for the crystal phonons are obviously based on the crystal space group,
or to be precise, on the corresponding factor group \cite{Turrell1972}. Of course, different $\kappa$-ET
salts may crystallize in different space group, so for instance \kHg crystal is monoclinic $C2/c$ ($C_{2h}^5$) \cite{Drichko2014}, with 4 (ET)$^+_2$ dimers in the unit cell (i.e. 2 dimers in the primitive cell), whereas $\kappa$-(ET)$_2$Cu[N(CN)$_2$]Br is orthorhombic, $Pnma$ ($D_{2h}^{16})$ with 4
dimers in the unit cell \cite{Maksimuk2001}. Therefore, in \kHg each dimer band is splitted into two, and in $\kappa$-(ET)$_2$Cu[N(CN)$_2$]Br it is splitted into four. However, since the four dimers
are arranged into two pairs located in different layers separated by the counterions, their interaction is expected to be negligibly small, and the four bands to be degenerate two by two \cite{Maksimuk2001}.
Here, I adopt the view that what it is important for the spectral predictions is the structure
of the 2D organic {\it layer} rather than the whole crystal, the arrangement of ET molecules
within the layer being the motif common to all $\kappa$ salts. Fig. \ref{fig:kappaetlayer}
show the typical chessboard packing of ET dimers within the layer, whose 2D symmetry
would be $Pgg$, to which the $C_{2h}$ factor group is associated. Thus, as shown from the right side of
Table I, each dimer phonon is coupled in-phase and out-of-phase, giving
rise to Raman active $A_g$ and $B_g$ Davydov components, and the corresponding
IR active $A_u$ and $B_u$ components.   

\subsection{Quantum chemical calculations}
\label{s:QCC}

Quantum chemical calculation have been performed with the
GAMESS package \cite{GAMESS} version 2022 R2, using DFT-B3LYP, and the 6-31G(d) basis
set, following the previously adopted protocol \cite{Girlando2011}.
The initial dimer geometry was taken from the 10 K structure of \kHg \cite{Drichko2014},
imposing the $D_2$ geometry for ET, and the inversion symmetry to (ET)$_2^+$ dimer, so the
charge is equally shared among the two molecules. In order to simulate the Raman polarization measurements, I had to post-process the GAMESS output to extract the
component of the Raman polarizability tensor along the $xyz$ coordinates of the dimer.
The oriented gas model \cite{Turrell1972} is then used to predict the $A_g$
and $B_g$ Raman polarization intensities (cf. Table I).
In oriented gas model the two dimers are assumed to be completely
non-interacting: the bands occur at the same frequency,
but the intensities are proportional to the square of the relevant polarizability component
along the crystal axes:
\begin{eqnarray} \label{eq:trans}
	\bm{\alpha}_{A_g} =& \bm{\mathrm{T}}\bm{\alpha}_{d1}\bm{\mathrm{T}}^\mathrm{t} +  
	\bm{\mathrm{T}}\bm{\alpha}_{d2}\bm{\mathrm{T}}^\mathrm{t} \\
	\bm{\alpha}_{{B}_g} =& \bm{\mathrm{T}}\bm{\alpha}_{d1}\bm{\mathrm{T}}^\mathrm{t} -  
	\bm{\mathrm{T}}\bm{\alpha}_{d2}\bm{\mathrm{T}}^\mathrm{t} 
\end{eqnarray}  
where $\bm{\alpha}_{d1},~\bm{\alpha}_{d2}$ are the polarizability tensors of dimer 1 and 2
(Fig. \ref{fig:kappaetlayer}),
expressed in the dimer $xyz$ coordinates, and $\mathbf{T}$ ($\mathbf{T}^t$)  are the
transformation matrix (its transpose) between the dimers' $xyz$ coordinates and the
crystal $abc$ coordinates (assumed orthogonal, since in \kHg the $\beta$ angle is very close to $90^\circ$
\cite{Drichko2014}). All intensities are {\it relative} intensities, as in the experiments.
Finally, in order to simulate the spectra of \kHg in the CO state \cite{Drichko2014,Hassan2020},
the constrain of the dimer inversion center has been removed, and an electric field
of 0.001 a.u. directed along the two molecular barycenters has been applied, in such a way to induce the required
charge separation. 

\subsection{Trimer model for inter-dimer e-mv interaction}
\label{s:trimermodel}
As stated above, the oriented gas model used to simulate the crystal Raman spectra
considers inter-dimer interactions as negligible, yielding at most a few wavenumber splitting.
This assumption is in general satisfied in molecular crystals, but in the present case
I have realized that e-mv interaction {\it between dimers}, rather than inside the dimer, originates quite sizable splitting clearly observable in Raman spectra.
The effect has been recognized in $\kappa$-(BEDT-TTF)$_2$Cu$_2$(CN)$_3$ \cite{Yakushi2015}, but not analyzed. To account for it I developed a suitable model along the lines proposed many years ago \cite{Rice1979,Yartsev1982,Painelli1986,Painelli1986a}
for e-mv interaction in CT crystals. The detailed development of the model is reported
in the Appendix, here the basic outline is presented.

At the core of traditional e-mv models there is the modulation of frontier molecular orbital
by the totally symmetric ($a_g$) molecular phonons. If the dimer is considered as a supermolecule,
disregarding the internal degrees of freedom, it is easy to realize that the in-phase coupled
$a_g$ modes ($\mathcal{A}_g$, cf. Table I) modulate the supermolecule
frontier orbital, and give rise to inter-dimer e-mv effects. In order to properly simulate
the Raman e-mv effects, a cluster made up of three dimers is used, in such a way to preserve
the inversion center between the dimers, as sketched in Fig. \ref{fig:kappaetlayer}.
The relevant cluster trimeric Hubbard Hamiltonian therefore is ($\hbar = 1,~i=1,3$):

\begin{eqnarray}
\label{eq:electronicH}
	H_e &=& \sum_{i}\varepsilon_0 n_i + \frac{U}{2}\sum_{i,\sigma}n_{i,\sigma}n_{i,-\sigma} + \\*
	 &+& t ~\sum_{\sigma}(c_{1,\sigma}^+ c_{2,\sigma}	+ c_{2,\sigma}^+ c_{1,\sigma} + c_{2,\sigma}^+ c_{3,\sigma} + c_{3,\sigma}^+ c_{2,\sigma}) \nonumber
\end{eqnarray}     
where $c_{i,\sigma}^+$ and $c_{i,\sigma}$ denote site $i$ creation and destruction operators 
for electron with spin $\sigma$,  $n_{i,\sigma} = c_{i,\sigma}^+c_{i,\sigma}$, $n_i = \sum_{\sigma} n_{i,\sigma}$ is the corresponding occupation number operator. $\varepsilon_0$ is the energy of the supermolecule frontier orbital, $U$ is the effective repulsion energy between two paired electrons on the same orbital, and $t$ is the inter-dimer hopping integral $t=t_{12}=t_{23}$. 

The above is the purely electronic Hamiltonian. The vibrational Hamiltonian is:
\begin{equation}
	H_v = \sum_{i,m} \frac{1}{4} (\dot Q_{i,m}^2+Q_{i,m}^2)\omega_m
\end{equation}
with $m$ counts the supermolecule $\mathcal{A}_g$ dimensionless normal coordinates $Q_m$.
Finally, the e-mv Hamiltonian is written as:
\begin{equation}
	H_{e-mv} = \sum_{m,i} g_m n_i Q_{i,m}
\end{equation}
where $g_m$ is the e-mv coupling constant, i.e., the derivative of the supermolecule
frontier orbital with respect to the relevant $\mathcal{A}_g$ normal coordinate.

The center of inversion symmetry is exploited by introducing the following number operators:
\begin{align}
	N_{\mathrm{TOT}}   &=  n_1+n_2+n_3~~~~~ = 3   \\
	\mathcal{N}_s  &=  2n_2 - (n_1+n_3)  \label{eq:Ns} \\
	\mathcal{N}_a  &=  n_1-n_3 
\end{align}
and symmetry adapted normal coordinates:
\begin{align}
	S_m &= Q_{1,m} + Q_{2,m} +Q_{3,m}  \\
    s_m & = 2Q_{2,m} - (Q_{1,m} +Q_{3,m}) \label{eq:s_m} \\
    q_m & = Q_{1,m} - Q_{3,m}
\end{align}

$N_{\mathrm{TOT}}$ is the total number of electrons distributed over the trimer cluster
and is a constant of motion. The corresponding coordinate $S_m$ is decoupled from the
electronic system. $\mathcal{N}_a$ and $q_m$ account for the effects of
inter-dimer e-mv perturbation in IR, and will not be discussed in the present paper, whereas
$\mathcal{N}_s$ and $s_m$ account for the e-mv effect in Raman.
The zero-frequency Raman susceptibility $\chi_R(0)$ is given by:
\begin{equation}
\chi_R(0)= \sum_{\mu} \frac{2\omega_{\mu 0} \langle \mu|\mathcal{N}_s|0\rangle ^2} {\omega_{\mu 0}}
\label{eq:chiR}
\end{equation}
where $\omega_{\mu 0} = E_\mu - E_0$ and $|\mu\rangle$ and $E_\mu$ are the eigenstates and eigenvalues
of the electronic Hamiltonian (\ref{eq:electronicH}), $\mu=0$ labeling the ground state. Finally,
in the isolated band approximation \cite{Pecile1989}, the frequency shift due to e-mv
perturbation is:
\begin{equation}
	\omega_m ^2 -\Omega_m ^2 = g_m ^2 \omega_m \chi_R(0)
	\label{eq:isolated_approx}
\end{equation}
where $\omega_m$ and $\Omega_m$ are the unperturbed and perturbed frequency of
the supermolecule $m$-th $\mathcal{A}_g$ phonon.

\section{Results}
\subsection{\kHg metallic state}
\label{s:metalic}
\begin{figure}
	\centering
	\includegraphics[width=0.9\linewidth]{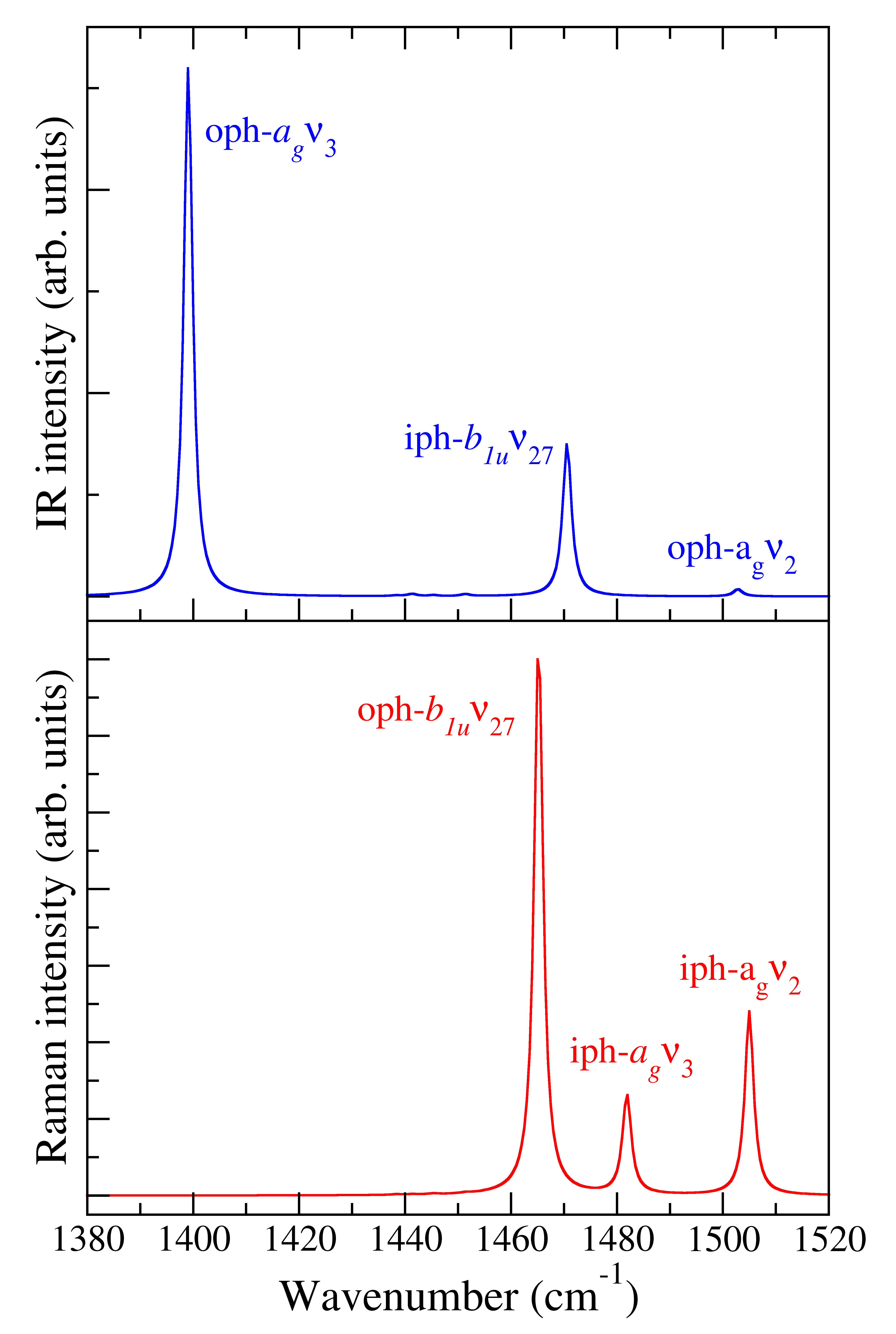}
	\caption{Calculated IR and Raman spectra of \kHg powders in the C=C stretching region. The C=C bands assignment refers to the conventional labeling relevant of a single ET molecule, the preceding oph and iph standing for out-of-phase and in-phase coupling of the molecular vibration in the ET dimer.}
	\label{fig:IRRampowderscalc}
\end{figure}

Fig. \ref{fig:IRRampowderscalc} shows the calculated total IR and Raman
intensities of (ET)$_2^+$ C=C stretching vibrations. They correspond to
the spectra of the  
powders of a typical $\kappa$-phase ET salt.
The chosen bandshape is a Lorentzian, with 1 \cm
bandwidth in order to have clearly separated peaks. Other modes, like the CH bendings,
are included in the simulation, but the spectra are dominated by six bands, three in IR and three in Raman, corresponding to the in-phase (iph) and out-of-phase (oph) coupling of the three vibrations depicted in Fig. \ref{fig:ET_C=Cvibs}. The simulated IR spectra have been already reported in Ref. \cite{Girlando2011}, here only a brief comment is in order. The iph-$b_{1u}~\nu_{27}$ is the charge
sensitive mode often used to determine the ET charge \cite{Girlando2011,Drichko2014}, whereas
the two oph modes are the e-mv induced vibrations, displaying frequency lowering with respect
to the corresponding iph modes, and intensity borrowing from the nearby CT electronic
transition \cite{Rice1979}. It is known that DFT-B3LYP calculations do not model very
well inter-molecular and CT interaction. The effects of e-mv interactions
are indeed only partially captured, as the calculated shifts and intensity borrowing are
highly underestimated: For instance, the experimental frequency of e-mv induced oph-$a_g~\nu_3$
is located around 1200 \cm in \kHg \cite{Drichko2014}, whereas the calculation places it around 1400 \cm. 

\begin{figure}
	\centering
	\includegraphics[width=0.98\linewidth]{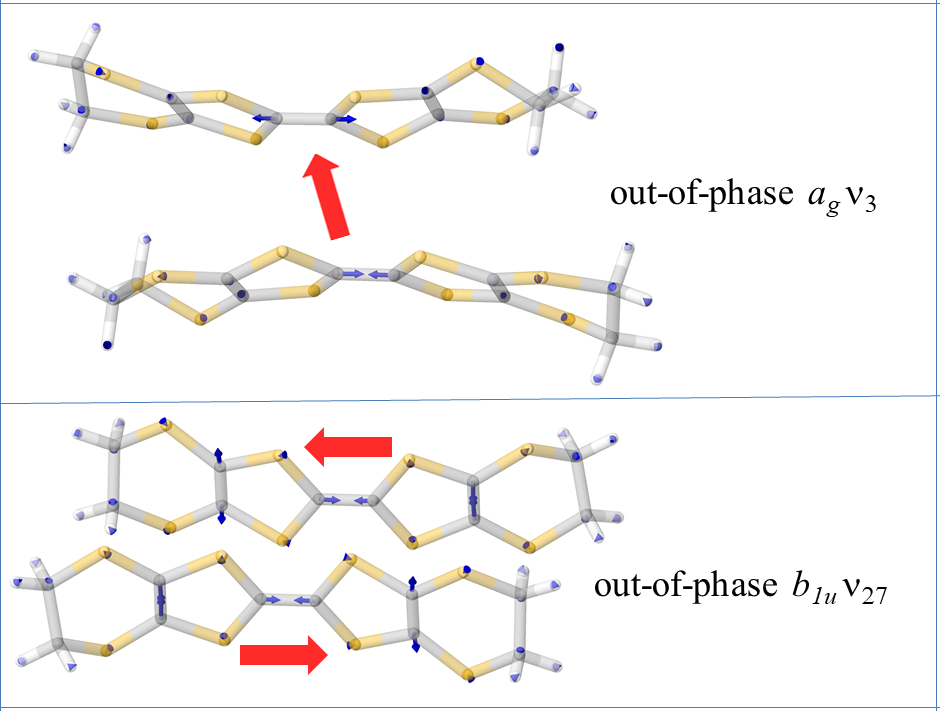}
	\caption{Top panel: Eigenvectors of the IR active, out-of-phase coupled $a_g~\nu_3$. Bottom panel: 	Eigenvectors of the Raman active, out-of-phase $b_{1u}~\nu_{27}$. The arrows indicate the
		direction of the oscillating dipole moments.}
	\label{fig:oph_nu3-nu27_eigenvectors}
\end{figure}

We notice that calculations predict that the intensity and frequency lowering of the
oph-$a_g~\nu_2$ are much lower than those of the $a_g~\nu_3$. This is due to the fact that the two modes are strongly mixed through the coupling to CT transition, 
and their behavior can be easily understood in terms of ordinary perturbation theory \cite{Painelli1986}. To simplify the following discussion, we shall focus attention on the only oph-$a_g~\nu_3$, whose eigenvectors are sketched in upper panel of Fig. \ref{fig:oph_nu3-nu27_eigenvectors}. The red arrow indicates the direction of the (strong) inter-molecular oscillating dipole moment, corresponding to back and forth charge flux \cite{Rice1979}.   

Now we analyze the simulated Raman spectra reported in the lower panel of Fig. \ref{fig:IRRampowderscalc},
where an unforeseen result is evidenced. In fact, three bands are present, the most intense being
associated to the out-of-phase coupling of the $b_{1u}~\nu_{27}$. Of course, selection rules
predict that this {\it ungerade} molecular vibration becomes Raman active in the dimer, but
its high intensity was not expected. And since the Raman spectra of $\kappa$-ET salts were dominated
by two bands, these were associated to the two ip-$a_g$ vibrations \cite{Maksimuk2001}, since totally symmetric modes generally are the most intense in the spectra. On the other hand, {\it a posteriori}
reasoning well accounts for the results of calculations: The $b_{1u} ~\nu_{27}$ has strong 
IR intensity, only surpassed by some e-mv induced modes, which means strong oscillating
dipole moment along the long ET molecular axis. And anti-phase oscillation of two parallel dipole
moments, as depicted in the bottom panel of Fig. \ref{fig:oph_nu3-nu27_eigenvectors}, means strong polarizability modulation, i.e. strong Raman intensity.

\begin{figure}
	\centering
	\includegraphics[width=0.85\linewidth]{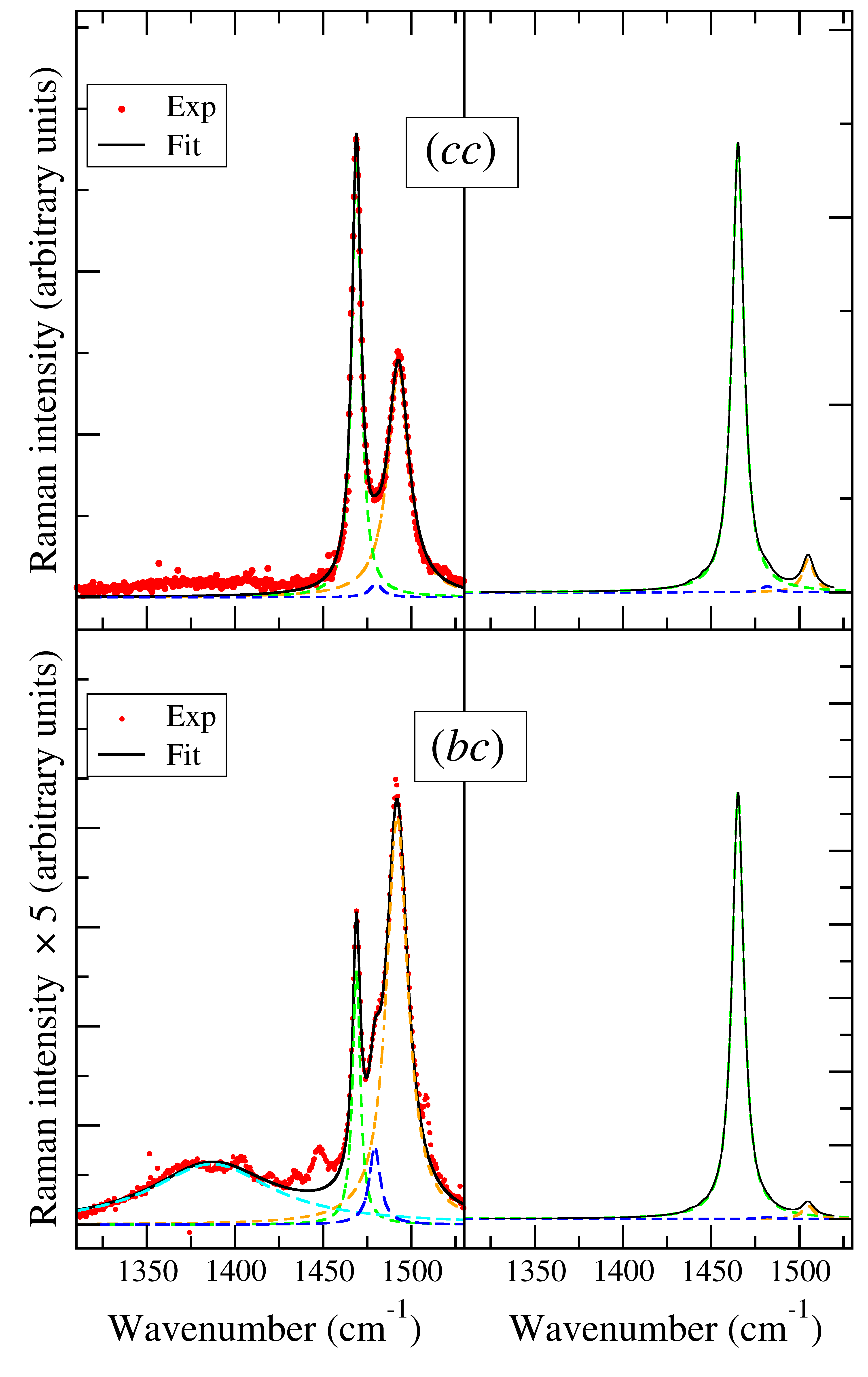}
	\caption{Left panels: \kHg experimental polarized Raman spectra at 50 K. Right panels: calculated spectra. The Raman intensities are in arbitrary units, but the scale of the both the $(bc)$ spectra is multiplied by a factor of $\approx$ 5.}
	\label{fig:ETHgCl_polarized}
\end{figure}

The assignment of the C=C Raman bands in ET $\kappa$ phase has to be reconsidered at the light of the above result. In Fig. \ref{fig:ETHgCl_polarized} we compare the experimental polarized Raman
spectra of \kHg on the $(bc)$ plane in the metallic phase \cite{Hassan2020} with the calculated one,
where 4 \cm bandwidth for the Lorentzians has been applied.  We first underline that the simulation reproduces properly the difference in intensity between the spectra recorded with parallel ($cc$) and crossed ($bc$) polarization, the latter being approximately four to five times less intense than the former.
The relative intensity within one polarization does not agree very well with experiment, but precise agreement is not to be expected. On one hand, calculations
are of course approximate, and do not consider the effects of resonance on the Raman
intensities, which is certainly present with the employed exciting lines \cite{Maksimuk2001}. On the other hand, the experimental Raman intensity measurements are delicate, with difficult to control effects related to resonance, crystal imperfections, etc. In any case the simulated spectra give 
an important information, namely that the ip-$a_g~\nu_3$
intensity is greatly reduced in the generally used collection geometry (backscattering observation on the most developed crystal plane, corresponding to the ET layers). Therefore the ip-$a_g~\nu_3$ is almost completely
masked by the nearby oph-$b_{1u}~\nu_{27}$, especially in the parallel or unpolarized spectra, so that only two Raman bands are generally observed in this spectral region.
The robustness of this results has been checked by changing DFT functional
and basis set. 

The precise interpretation of the crossed polarization, ($bc$) spectra of \kHg is
more complicate, as several bands are clustered in the 1450-1520 \cm spectral region. 
In addition, there is a fourth band, centered around 1387 \cm, very broad,
but with a total intensity comparable to that of the other bands. This peculiar Raman band is found in other $\kappa$-phase salts, like for instance $\kappa$-(BEDT–TTF)$_2$Cu(NCS)$_2$ \cite{Revelli2021} or $\kappa$-(BEDT-TTF)$_2$Cu[N(CN)$_2$]Br \cite{Maksimuk2001}, albeit at
frequencies different in different salts.
Isotopic substitution of the central C=C carbon atoms \cite{Maksimuk2001}
shows that this band is associated to the crystalline $B_g$ component corresponding to the $a_g~\nu_3$ molecular mode (Table I). The question now is: Why this component displays such an unusually large splitting with respect to $A_g$ component ? Below I shall show that the effect is a manifestation of {\it inter}-dimer e-mv coupling that has not been properly analyzed so far.

 \begin{figure}
 	\centering
 	\includegraphics[width=0.8\linewidth]{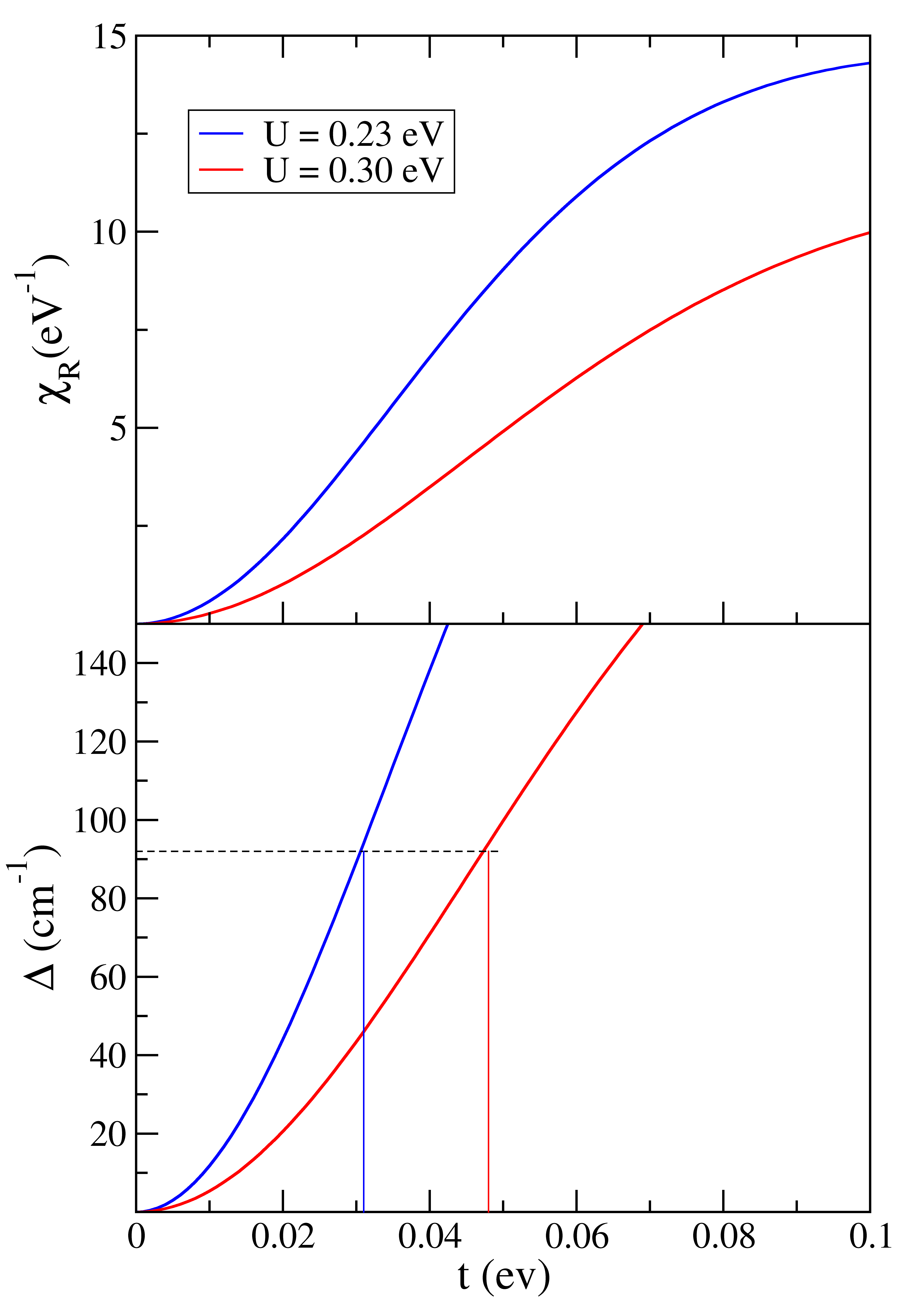}
 	\caption{Results of the trimer model. Top panel: $\chi_R(t)$ for two different values of $U$; Bottom panel: $t$ dependence of $\Delta = \omega_{\nu3} - \Omega_{\nu3}$. The dashed horizontal line
 		marks the position of the experimental $\Delta = 92$ \cm for \kHg at 50 K (from left bottom panel of Fig. \ref{fig:ETHgCl_polarized}).}  
 	\label{fig:chidelta_t}
\end{figure}
  
 As already explained in Section \ref{s:trimermodel} and depicted in Fig. \ref{fig:kappaetlayer},
 in order to deal with inter-dimer e-mv interaction a trimer model is adopted, the trimer being made
 up by (ET)$_2^+$ dimers.  Appendix \ref{appendixA} reports the solution of the model:
 The electronic Hamiltonian of Eq. \ref{eq:electronicH} is solved numerically for different
 values of the $U/t$ ratio, and so is the Raman susceptibility, whose $t$ dependence for
 two different values of $U$ is reported in the top panel of Fig. \ref{fig:chidelta_t}.
 We then use Equation \ref{eq:isolated_approx} with $g_{\nu3}$ = 71 meV \cite{Girlando2011}   
 to calculate $\Delta = \omega_{\nu3} - \Omega_{\nu3}$, as reported in the bottom panel of
 Fig. \ref{fig:chidelta_t}. By taking the experimental $\Delta = 92$ \cm we get
 $t = 31$ meV for $U=0.23$ eV or $t = 48$ meV for $U= 0.30$ eV, values compatible (considering
 all the made approximations) with those calculated in the framework of the
 effective dimer model \cite{Gati2018}.
 
The effective dimer model \cite{Tamura1991,Kino1995,McKenzie1999} is often used to
analyze the phase diagram of $\kappa$-phase ET salts. Indeed, the two parameters of the
model Hamiltonian (\ref{eq:electronicH})  correspond to two of the four parameters of the
effective dimer model (see Appendix \ref{appendixA}). The work here, making reference to just one experimental datum, can only give the ratio between the two parameters, and here we shall not go beyond the demonstration of the e-mv origin of the 1479 \cm Raman band. But the discovery of the effect
opens the way to the possible experimental estimate of all the $\kappa-$phase effective dimer model
parameters {\it via} optical spectroscopy.

\subsection{\kHg charge ordered state}
\label{s:CO state}

\begin{figure}[ht]
	\centering
	\includegraphics[width=0.9\linewidth]{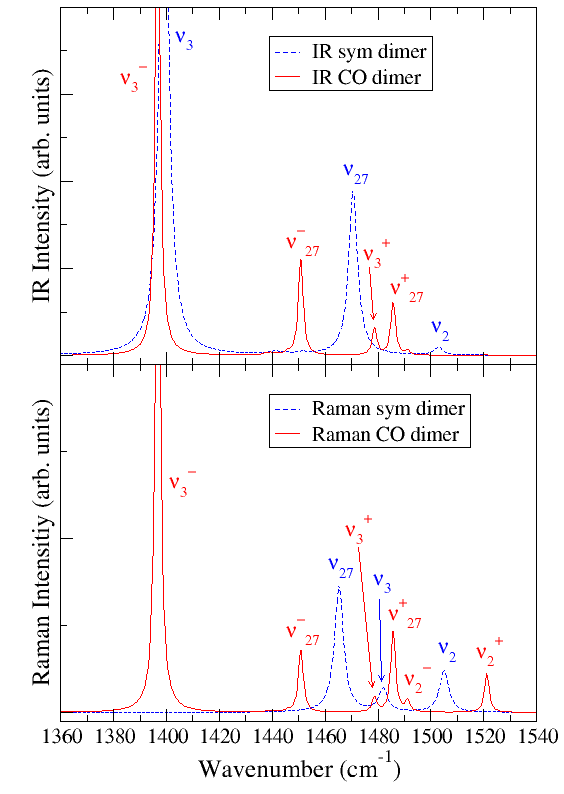}
	\caption{Calculated IR and Raman spectra of \kHg powders in the C=C stretching region above and below a hypothetical CO phase transition, i.e. a symmetric dimer (dashed blue line) and a CO dimer (red line). The mode labeling is the same as in Fig. 3, and in the CO phase the plus and minus refer to charge rich and charge poor molecule.}
	\label{fig:COpowderscalc}
\end{figure}

Among $\kappa$-phase salts, \kHg is unique in displaying a CO phase transition around 30 K, as detected
by both IR \cite{Drichko2014} and Raman \cite{Hassan2020} spectroscopy. With the
aim of shedding light into the transition mechanism and physical
consequences, I decided to simulate
the vibrational spectra of \kHg in the CO state, following the generally
accepted idea that the CO or disproportionation  occurs inside the dimer.
Therefore I removed the constraint of the inversion
center in the dimer, that assured equally distributed charge between the
two ET moieties, at the same applying an electric field along the molecular barycenters to induce the experimentally observed 0.2$e$ charge difference
(cf. Section \ref{s:QCC}).
Spectral predictions
say that by removing the inversion center all the modes becomes both IR and Raman active, so,
disregarding the Davydov splitting, we expect six C=C bands occurring at the same frequency
in IR and Raman. Of course, this is precisely what we obtain in the simulation, but with
the advantage of information about the intensity distribution, as shown in Fig. \ref{fig:COpowderscalc}, where
the simulated spectra of the powders are reported 
for both the symmetric and CO dimer.

The band labeling of Fig. \ref{fig:COpowderscalc} is the the same as Fig. \ref{fig:IRRampowderscalc},
with omission of the `oph' and `iph' prefixes to avoid label crowding. In the CO phase (red line), the
`$+$' and `$-$' superscripts refer to charge rich and charge poor
molecule within the dimer. We first examine the IR spectra in the top panel of the Figure.
Here we see that the calculated 35 \cm splitting of the charge sensitive $\nu_{27}$ is in line
with the experimental one (29 \cm \cite{Drichko2014}), indicating that the simulation correctly
reproduces the actual charge separation. The $\nu_{27}$ splitting is observed in the polarization
perpendicular to the ET layer, i.e. the $bc$ plane. On the other hand, the experimental spectra along $b$ and $c$ are dominated by the e-mv coupled mode $\nu_3$ and the nearby CT electronic transition, and become very complicated below the phase transition \cite{Drichko2014}. The calculation indicates that $\nu_3^-$ is downshifted only by a few wavenumbers with respect of the $\nu_3$ of the metallic phase (dashed blue line in the Figure), while the $\nu_3^+$, whose intensity is predicted to be much lower, is close to the
almost undetectable $\nu_2^\pm$. We have already remarked in Section
\ref{s:metalic} that the calculated frequency of the $\nu_3$
($\sim 1400$ \cm) is much higher than the observed one (1200 \cm) \cite{Drichko2014}. 
 
 In the simulated Raman spectra (bottom panel of Fig. \ref{fig:COpowderscalc}) we again
 notice that the splitting of the charge sensitive $\nu_2$ mode, 30 \cm, is in line with
 the experiment (32 \cm \cite{Hassan2020}), whereas the remaining
 1440-1500 \cm spectra region is very
 crowded. Before embarking in the comparison with the corresponding experimental
 spectra, it is useful to focus on a very important and unambiguous information
 provided by the simulation,   
 namely, the appearance with huge intensity the e-mv coupled, out-of-phase $a_g \nu^-_3$
 (top panel of Fig. \ref{fig:oph_nu3-nu27_eigenvectors}). 
 Elementary group theory just predicts that the all the
 IR active modes, forbidden in the symmetric dimer
 (metallic phase) become also Raman active if the
 dimer inversion center is lost.  Perhaps the huge Raman intensity
 is overestimated in a calculation made in the presence of an external electric field, but after all this finding parallels the
 high intensity of the out-of-phase $b_{1u}~\nu_{27}$: In
 both cases we have strong charge oscillation with corresponding
 oscillating dipole moment, reflecting in equally strong oscillating
 polarizability.
 
\begin{figure}
 	\centering
 	\includegraphics[width=0.9\linewidth]{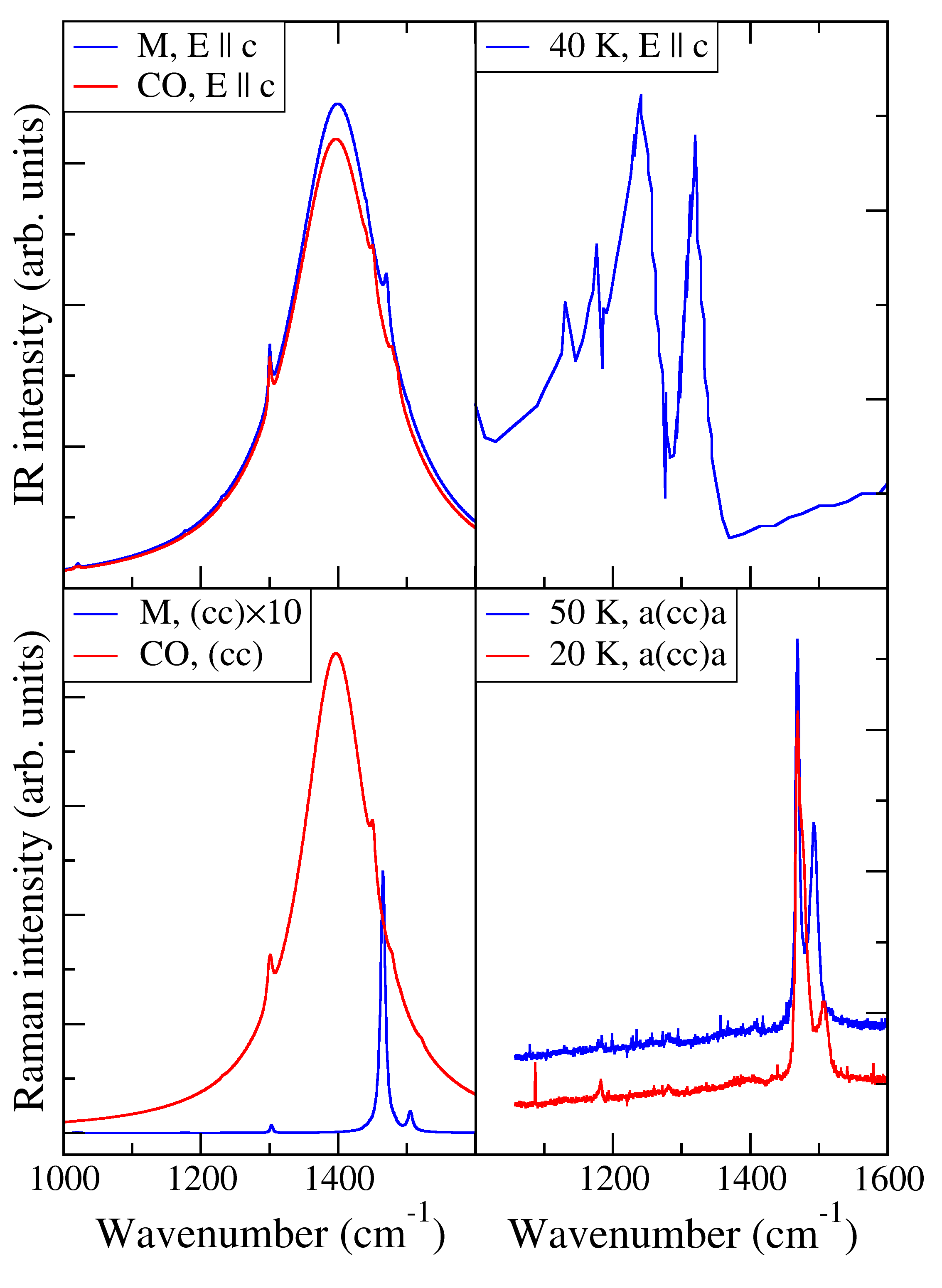}
 	\caption{Left panels: Simulated and experimental polarized Raman and IR spectra of \kHg in the metallic (M) and CO phases (blue and
 		red lines. respectively). The simulated spectra (left panels) assume charge ordered, non symmetric dimers.}
 	\label{fig:ethgcl-cocryst}
 \end{figure}

Therefore according to the simulation, if \kHg CO transition
involves differently charged molecules \textit{within} the dimer,
we should observe the insurgence of a strong Raman band
in correspondence with the IR e-mv induced $a_g~\nu_3$.
In order to verify the above, in Fig. \ref{fig:ethgcl-cocryst} we compare the simulated crystal Raman spectra of symmetric and CO dimers with the experimental ones above and below the phase transition. We have chosen the ($cc$) polarization, where the out-of-phase $a_g \nu_3$
should appears with the highest intensity.
In the simulated spectra (left side of Fig. \ref{fig:ethgcl-cocryst})
we have used Lorentzian bandshape with bandwidth adjusted to
simulate the experiment.
 
The upper right panel of Fig. \ref{fig:ethgcl-cocryst} reports the
experimental conductivity spectrum in the metallic phase (digitized from Ref. \cite{Drichko2014})
The one in the CO state is not reported since at the phase
transition the reflectivity goes down drastically and the
optical conductivity spectral weight shifts from the zero-frequency peak to a new maximum at about 700 \cm \cite{Drichko2014}, overlapping the $a_g \nu_3$.  But its position is substantially
unaltered, as confirmed by the simulated spectra. Finally we recall
that the experimental IR bandshape of the $a_g \nu_3$ is not a Lorentzian, but it has rather a Fano-like aspect, with Fano indentations in correspondence with the underlying CH bending modes \cite{Sedlmeier2012}.   

\begin{figure}
	\centering
	\includegraphics[width=0.8\linewidth]{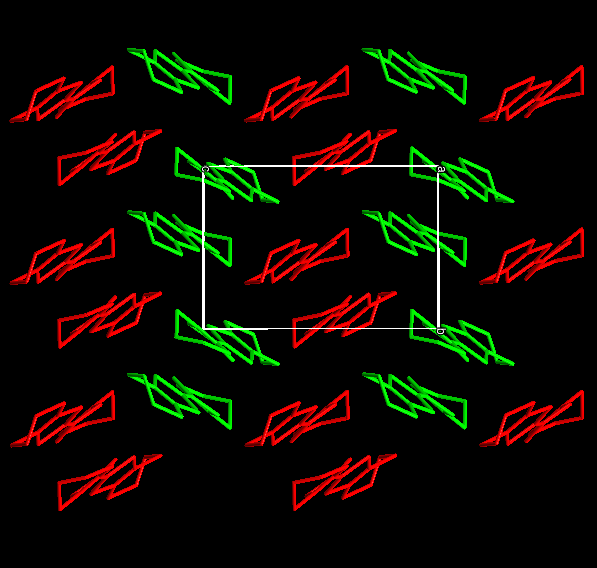}
	\caption{Suggested crystal structure of \kHg in the CO phase. Symmetry (inversion center) related dimers are drawn in the same color, red or green. The red and green color correspond to
		differently charged dimers. Hydrogens are not shown.}
	\label{fig:COstruct_pattern}
\end{figure}

The right bottom panel of Fig. \ref{fig:ethgcl-cocryst} shows 
that the experimental Raman spectra in the frequency region 
of the out-of phase $a_g \nu_3$ are completely flat  
above the phase transition, at it should be, but remain precisely
the same below. We are then led to the inescapable conclusion
that the intra-dimer inversion center is retained in the CO phase.
In other words, and contrarily to assumption at the basis of the present simulation,
the charge separation is
\textit{not} between the two moieties of the dimer, but between
the dimers, as sketched in Fig. \ref{fig:COstruct_pattern}. The actual crystal structure should be $P\overline{1}$, in place of the average structure $C2/c$ detected at the synchrotron \cite{Drichko2014}.

The above indication contrasts with
the intra-dimer disproportionation proposed early on the
basis of electronic and magnetic data \cite{Drichko2014,Gati2018,Hassan2020}.
On the other hand the present finding is supported by what it has been found
on another $\kappa$-phase salt, the widely studied spin-liquid candidate $\kappa$-(BEDT-TTF)$_2$Cu$_2$(CN)$_3$\cite{Pustogow2022}. A very accurate crystal structure determination \cite{FouryLeylekian2018} has shown a small {\it inter}-dimer average charge difference, $0.06 \pm 0.05$, and according a recent paper \cite{Liebman2024} no strong Raman band seems to be present around 1200 \cm,
in correspondence with the IR e-mv induced $a_g \nu_3$. 
Another intuitive argument in favor of the inter-dimer CO is that the itinerant electrons jump
more easily and quickly between more strongly interacting molecules, and in the present case
the intra-dimer hopping integral is about twice the inter-dimer one \cite{Gati2018}.

Of course it is impossible to make a computational simulation of the Raman spectra
of a inter-dimer charge-ordered $\kappa$-phase to compare with the experiment.
A tentative interpretation of \kHg Raman spectra in CO state based on the assumption that the spectra are a superposition of (BEDT-TTF)$_2^{+0.6}$ and 
(BEDT-TTF)$_2^{+0.4}$ symmetric dimers is reported in Appendix \ref{appendixB}.

\section{Conclusions}
First principles calculations of the Raman intensities of (BEDT-TTF)$_2^+$
dimers oriented as in \kHg crystal has lead us to reconsider
the assignment of the corresponding Raman spectra in the critical region
of BEDT-TTF C=C stretching. Although group theory is predicting that
in a dimer the out-of-phase coupling of the IR active $b_{1u}~\nu_{27}$
becomes Raman active, no one expected that the corresponding band
had such a strong Raman intensity, likely obscuring the in-phase coupling
of the $a_g~\nu_3$ (cf. Figs. \ref{fig:IRRampowderscalc} and \ref{fig:ETHgCl_polarized}). An intuitive
justification of this finding is based on the fact that the $b_{1u}~\nu_{27}$
implies large charge oscillation in the direction of the BEDT-TTF
long molecular axis \cite{Girlando2011}, leading to strong Raman
intensity for the corresponding anti-phase charge oscillation
(bottom panel of Fig. \ref{fig:oph_nu3-nu27_eigenvectors}).

In the reanalysis of the C=C stretching Raman spectra,
we have also been able identify the origin of a broad
band appearing in the cross polarized spectra about
90 \cm lower than the $a_g~\nu_3$. This band is typical
of the $\kappa$-phase salts, and it has been already
assigned to the $B_g$ component of $A_g$ phonon associated
to $a_g~\nu_3$ stretching \cite{Maksimuk2001}. We explain
the unusually large Davydov splitting as a manifestation
of a so far unrecognized inter-dimer e-mv coupling.
By adapting the Rice's model of e-mv coupling \cite{Rice1979}
to the present case, we show how optical data can be
used to extract relevant parameters of the effective dimer model
\cite{Tamura1991,Kino1995,McKenzie1999} often adopted to explain the physical properties of $\kappa$-phase ET salts. 

Finally this paper reports the simulated Raman spectra of the CO
ferroelectric phase of \kHg \cite{Drichko2014,Hassan2020}
by assuming that the CO implied the breaking of the intra-dimer
inversion center symmetry. In such a case it is obvious
that the IR active phonons become also Raman active and
vice-versa, as observed in another quite different
ET  salt, $\theta$-(BEDT-TTF)$_2$TlZn(SCN)$_4$ \cite{Suzuki_2004}. What the calculations do again tell us, however, is the huge Raman intensity acquired by the out-of-phase
coupling of the $a_g~\nu_3$. This phonon has huge IR intensity
due to the inter-molecular charge oscillation induced by the
e-mv coupling \cite{Rice1979} (cf. the top panel of Fig.  \ref{fig:oph_nu3-nu27_eigenvectors}), and as in the case of $b_{1u}~\nu_{27}$, should give rise to equally large Raman intensity.
The lack of the appearance of such a band in correspondence with \kHg CO phase transition (cf. Fig.\ref{fig:ethgcl-cocryst}) implies that
the intra-dimer inversion center symmetry is not broken,
strongly suggesting that the CO pattern (Fig. \ref{fig:COstruct_pattern})
is different from the one hypothesized so far \cite{Drichko2014}.
The above finding is likely the most important obtained here, and I believe
it will prompt the Raman collection and analysis of other 
$\kappa$-BEDT-TTF and BETS salts (BETS is bis-ethylenedithio-tetraselenafulvalene),
that constitute a widely investigated family of frustrated Mott systems \cite{Riedl2022}.

\section*{Acknowledgements}
I gratefully thank Prof. N. Drichko (John Hopkins Baltimore), Prof. M. Dressel (Stuttgart University) and Prof. M. Masino and A. Painelli (Parma University)
for the many useful discussions and critical reading of the manuscript.
I also thank N. Drichko for calling my attention to the problem and providing the original and in part unpublished spectra of \kHg.


\appendix
\counterwithin{figure}{section}
\counterwithin{table}{section}
\section{Trimer model for interdimer e-mv coupling}
\label{appendixA}

The first step is to solve the electronic hamilonian (\ref{eq:electronicH}). The basis set
is constructed by distributing three spins on the three dimer sites. There are
20 ways of distributing the spins, 4 constitute a quartet ($S=3/2$) and 16 are doublets ($S=1/2$),
8 states correspond to different spin arrangements with one electron per site, as
summarized in Table A.1.

\begin{table}
	\label{tab:e-distribution}
\begin{center}
\caption{Possible distribution of three electrons on three sites}
\vskip 0.2 cm
\begin{tabular}{|c | c c c|c|c|}
	\hline
 n	& site 1 & site 2 & site 3 & $S_z$ & wavefunction \\
	\hline \hline
1	& $\uparrow$ & $\uparrow$ & $\uparrow$ & 3/2 &  \\
2	& $\downarrow$ &  $\downarrow$ &  $\downarrow$ & -3/2 &  \\
3	& $\uparrow$ &  $\uparrow$ & $\downarrow$  & 1/2 &  \\
4	& $\downarrow$ & $\downarrow$ & $\uparrow$  & -1/2 &  \\
\textcolor{red}{5}	& $\uparrow$ & $\downarrow$ & $\uparrow$ & 1/2 & \textcolor{red}{$\phi_1$}  \\
6	& $\downarrow$ & $\uparrow$ & $\downarrow$ & -1/2  &  \\
7	& $\downarrow$ & $\uparrow$ & $\uparrow$ & 1/2 &  \\
8	& $\uparrow$   & $\downarrow$ & $\downarrow$ & -1/2 &   \\
\hline
\textcolor{red}{9}	& $\uparrow\downarrow$  & 0 & $\uparrow$ & 1/2 & \textcolor{red}{$\phi_2$} \\
10	& $\uparrow\downarrow$ & 0 & $\downarrow$ & -1/2 &  \\
\hline
\textcolor{red}{11}	& $\uparrow\downarrow$ & $\uparrow$  & 0 & 1/2 & \textcolor{red}{$\phi_3$} \\
12	& $\uparrow\downarrow$ & $\downarrow$ & 0 & -1/2 &  \\
\hline
\textcolor{red}{13}	& $\uparrow$ & $\uparrow\downarrow$  &  0 & 1/2 & \textcolor{red}{$\phi_4$} \\
14	& $\downarrow$ & $\uparrow\downarrow$ & 0 & -1/2 &  \\
\hline
\textcolor{red}{15}	& 0 & $\uparrow\downarrow$ & $\uparrow$ & 1/2 & \textcolor{red}{$\phi_5$} \\
16	& 0 & $\uparrow\downarrow$ & $\downarrow$ & -1/2 &  \\
\hline
\textcolor{red}{17}	& 0 & $\uparrow$ & $\uparrow\downarrow$ & 1/2 & \textcolor{red}{$\phi_6$} \\
18	& 0 & $\downarrow$ & $\uparrow\downarrow$ & -1/2 &  \\
\hline
\textcolor{red}{19}	& $\uparrow$ & 0 & $\uparrow\downarrow$ & 1/2 & \textcolor{red}{$\phi_7$} \\
20	& $\downarrow$ & 0 & $\uparrow\downarrow$ & -1/2 &  \\
	\hline \hline
\end{tabular}
\end{center}
\end{table}

Since we are not concerned with the spin, but only with the distribution of the three charges
following the Pauli exclusion principle, we can take as basis functions just the ones with the same
$S_z = 1/2$ (marked in red in the Table). In this way 
the basis function set is reduced to seven, $\phi_1$ to $\phi_7$.   
Identical result can be obtained in simpler way by using a spinless 
fermion approach. Adopting the symbols
$\Circle,~ \logof$ and $\CIRCLE$ for empty site,
single and double occupancy, respectively,
the resulting seven basis wave functions can be written as:
$\phi_1 = |\logof\logof\logof \rangle$, 
~$\phi_2 = |\CIRCLE~\Circle \logof \rangle$, 
~$\phi_3 = |\CIRCLE \logof \Circle \rangle$, 
~$\phi_4 = |\logof \CIRCLE~\Circle \rangle$, 
$\phi_5 = |\Circle~ \CIRCLE \logof \rangle$, 
$\phi_6 = |\Circle \logof \CIRCLE \rangle$, and 
$\phi_7 = |\logof  \Circle~ \CIRCLE \rangle$.

We can exploit the center of
inversion to construct symmetry adapted wavefunctions:
\begin{eqnarray}
	\psi_1^{(+)} & \equiv & \phi_1 \nonumber \\
	\psi_2^{(+)}   & = & (1/\sqrt{2}) ~\{\phi_3 + \phi_6\} \nonumber \\
	\psi_3^{(+)}  & = & (1/2) ~ \{[\phi_5 + \phi_4] + [\phi_2 + \phi_7]\}  \nonumber \\ \nonumber \\
	\psi_4^{(\mp)} & = & (1/2) ~\{[\phi_5 + \phi_4] - [\phi_2 + \phi_7]\} \\  \nonumber \\
	\psi_5^{(\pm)} & = & (1/2) ~\{[\phi_5 - \phi_4] + [\phi_2 - \phi _7]\}  \nonumber \\ \nonumber \\
	\psi_6^{(-)}   & = & (1/\sqrt{2}) ~ \{\phi_3 - \phi_6\} \nonumber \\
	\psi_7^{(-)}   & = & (1/2) ~ \{[\phi_5 - \phi_4] - [\phi_2 - \phi _7]\} \nonumber 
\end{eqnarray}  

\noindent
On this basis the electronic Hamiltonian (\ref{eq:electronicH}) is block diagonal ($\varepsilon_0 = 0$):
\begin{equation}
\begin{bmatrix}
	0  &   & 2t         & 0 & 0 & 0 & 0 \\
	0  &  U  & \sqrt{2}t  & 0 & 0 & 0 & 0 \\
	2t & \sqrt{2}t    & U & 0 & 0 & 0 & 0 \\
	0  &   0 & 0 & U & 0 & 0 & 0 \\
	0  &   0 & 0 & 0 & U & 0 & 0 \\
	0  &  0  & 0  & 0 & 0 & U & -\sqrt{2} t \\
	0  &  0  & 0  & 0 & 0 & -\sqrt{2}t & U \\  
\end{bmatrix}
\end{equation}

\begin{figure}
	\centering
	\includegraphics[width=0.9\linewidth]{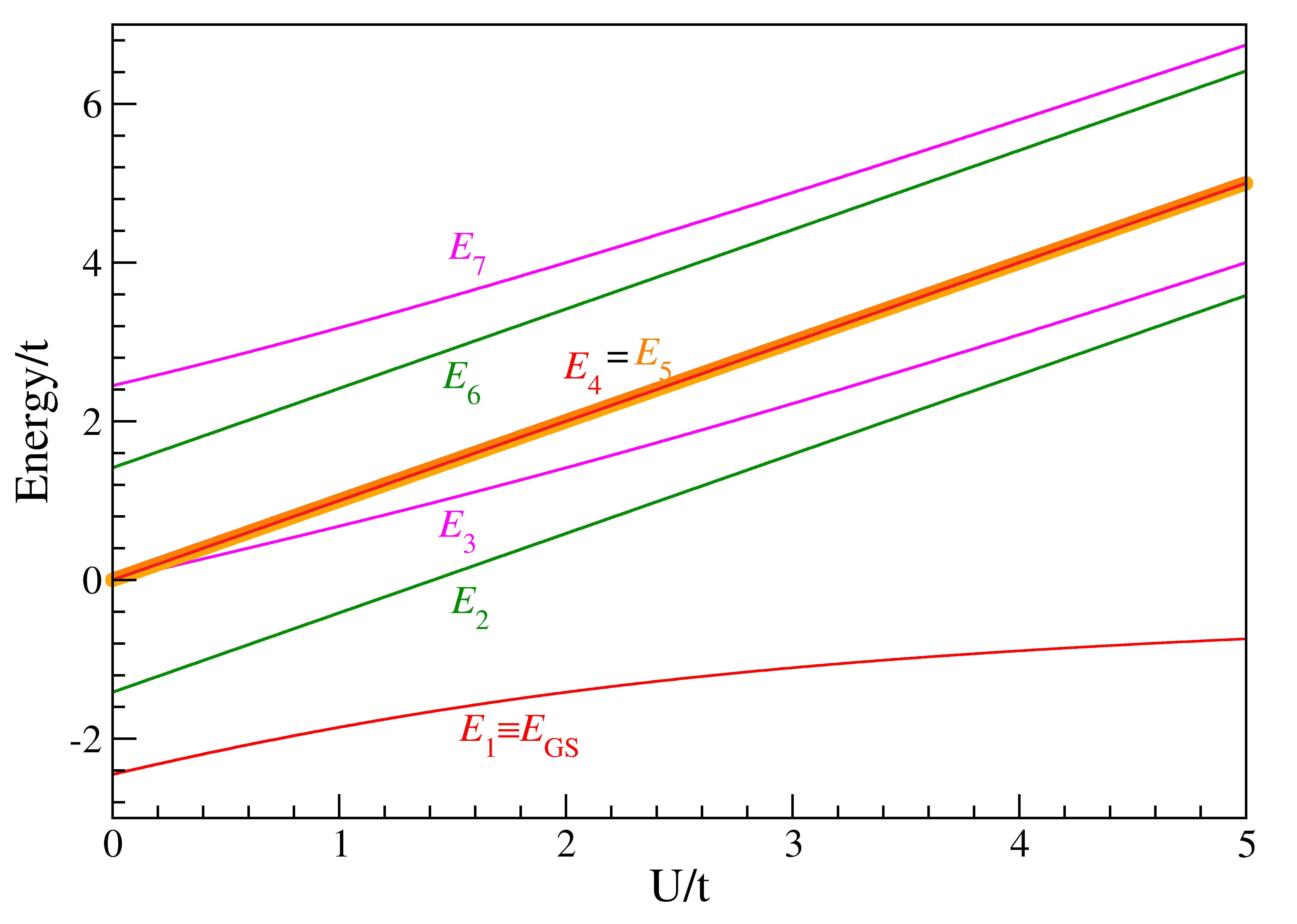}
	\caption{Trimer energy levels as function of $U$, in $t$ units ($\varepsilon_0=0$)}
	\label{fig:e-vsu}
\end{figure}

The $3\times 3$ block can be easily diagonalized symbolically only for $U=0$, with roots $E_1 \equiv E_{\mathrm{GS}} = - \sqrt{6}t$, $E_3 = 0$ and $E_7=\sqrt{6}t$. The $2 \times 2$ block gives $E_2 = U - \sqrt{2} t$ and $E_6 = U +\sqrt{2}t$, while the $E_4 = E_5= U$. The numerical solution for the 
energies as a function of $U$ for $t=1$ are reported in Fig. \ref{fig:e-vsu}.

The ground state wavefunction is:
\begin{equation}
	\Psi_{\mathrm{GS}} \equiv \Psi_1 = c_1 	\psi_1^{(+)} + c_2 \psi_2^{(+)} + c_3 \psi_3^{(+)} 
\end{equation}

\noindent
and the operator (\ref{eq:Ns}) connects the ground state with $\Psi_4 \equiv \psi_4^{(\mp)}$ only, so that the Raman susceptibility (\ref{eq:chiR}) is given by:
\begin{equation}
	\chi_R = \frac{2\langle \Psi_4|\mathcal{N}_s|\Psi_{\mathrm{GS}}\rangle^2}{U-E_\mathrm{GS}} = \frac{18 c_3^2}{U-E_\mathrm{GS}}
\end{equation}
The top panel of Fig.\ref{fig:chidelta_t} reports $\chi_R (t)$ for two different values of $U$.

\begin{figure}
	\centering
	\includegraphics[width=0.6\linewidth]{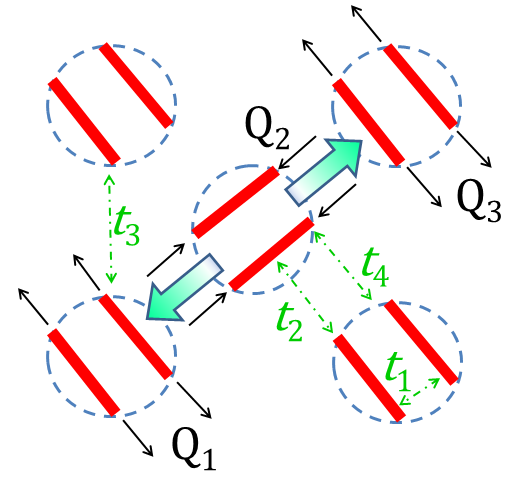}
	\caption{Cartoon of the e-mv mechanism yielding to the observed Raman effect.
	The definition of the four inter-molecular hopping integrals $t_1 - t_4$ is also shown.
    Within the effective dimer model $U_{\textrm{eff}} = 2t_1$, $t=(t_2+t_4)/2$, $t^\prime = t_3/2$.}
	\label{fig:emvcartoon}
\end{figure}

In Fig. \ref{fig:emvcartoon} we show a cartoon with an intuitive picture of the interdimer e-mv
mechanism. $Q_1$, $Q_2$ ans $Q_3$ are the normal coordinates (relative to iph-$a_g~\nu_3$) that 
combined in anti-phase ($B_g$, Table I) yield the $s_m$ coordinate (Eq. \ref{eq:s_m}). This coordinate, modulating the dimers frontier orbitals in anti-phase, is coupled to the symmetric inter-dimer CT, as sketched by the greenish arrows, thus lowering the corresponding vibrational frequency.

In the same Figure we report the definition of the inter-molecular hopping integrals $t_1$ (between the two molecules in the dimer), $t_2$, $t_3$ and $t_4$. In the framework of the effective dimer model
\cite{Tamura1991,Kino1995,McKenzie1999}
we have $U_{\textrm{eff}} = 2t_1$, $t=(t_2+t_4)/2$, $t^\prime = t_3/2$. Obviously, $U_{\textrm{eff}}$ and
$t$ coincide with $U$ and $t$ of our electronic Hamiltonian (\ref{eq:electronicH}).

\counterwithin{figure}{section}
\counterwithin{table}{section}
\section{\kHg CO phase: interpretation of the Raman spectra in the C=C spectral region.}
\label{appendixB}

\begin{figure}
	\centering
	\includegraphics[width=0.9\linewidth]{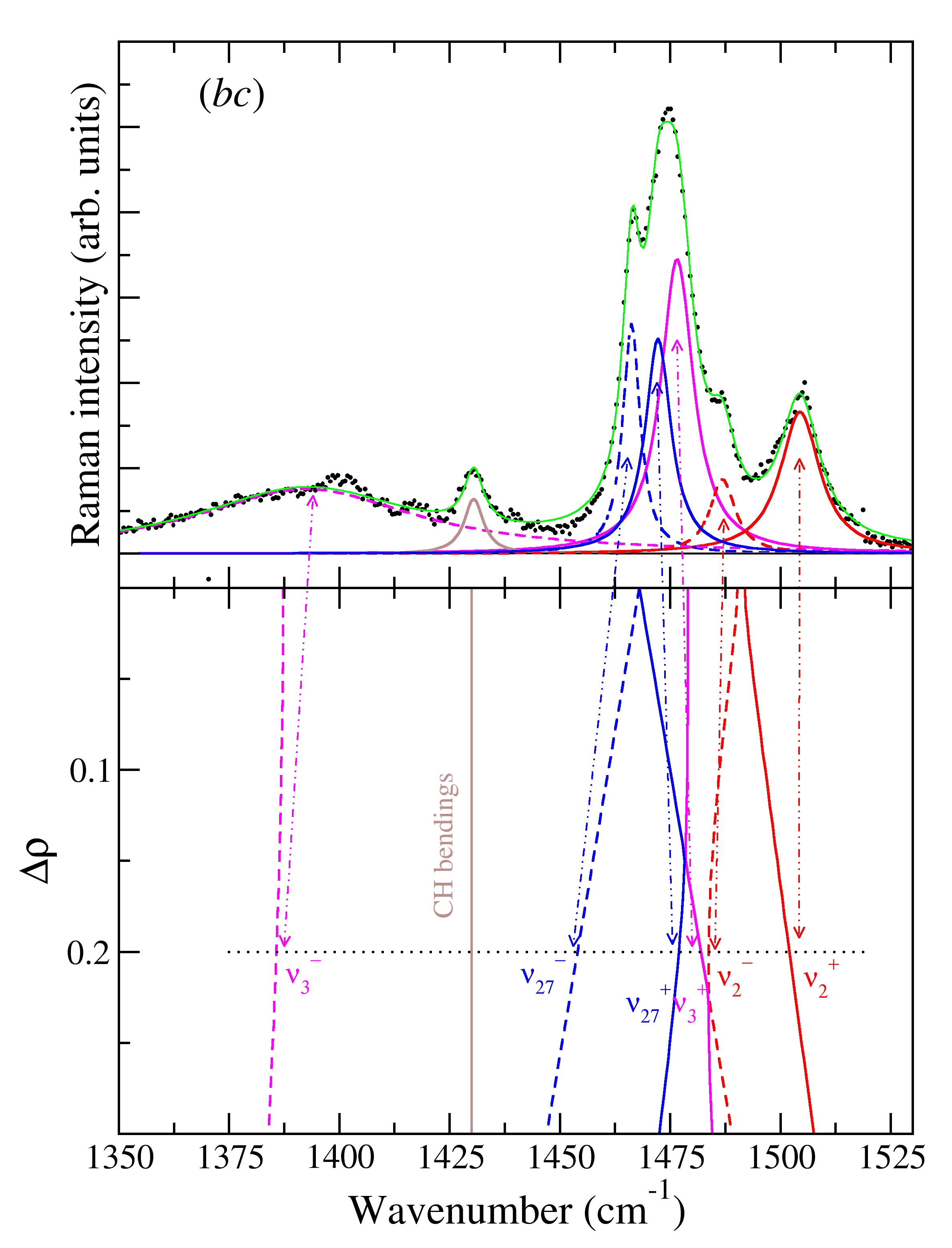}
	\caption{Top panel: Experimental Raman spectrum ($bc$ polarization) in the CO phase of \kHg (from Ref. \cite{Hassan2020}). Bottom panel: $\Delta\rho$ dependence of the Raman frequencies, taking into account the effect of inter-dimer e-mv interaction. The double arrows connecting the two panels
		indicate the proposed assignment of the bands. The horizontal dotted line
		corresponds to the observed $\Delta\rho = 0.2$.}
	\label{fig:S_RamanCO}
\end{figure}

The interpretation of the Raman spectra of \kHg in the CO phase is difficult, as many
bands crowd in the C=C spectral region. This is particularly evident in the
($bc$) polarized spectrum (top panel of Fig. \ref{fig:S_RamanCO}), whereas the parallel polarized spectra (not shown) are dominated a very intense band obscuring the underlying structure.

According the discussion in Section \ref{s:CO state}, the CO pattern
is made up of differently charged (+0.4 and +0.6 $e$) ET dimers (Fig \ref{fig:COstruct_pattern}), and of
course the corresponding Raman spectrum cannot be simulated by calculation.
The spectral interpretation here is based on a phenomenological
approach aimed at reproducing and assigning the experimental frequencies.
The frequencies of three C=C stretching (iph-$a_g \nu_2$ and $\nu_3$, oph-$b_{1u} \nu_{27}$) are known to depend linearly from the charge $\rho$
according to \cite{Girlando2011}:
\begin{equation}
	\omega_i^\pm(\Delta\rho) = \omega_i(0.5) \pm \frac{\Delta\rho \cdot \varDelta_i}{2} 
\end{equation}
where the mode frequency $\omega_i^\pm$ of the charge rich
and charge poor ET dimer is referenced to the experimental frequency at $\rho = 0.5$ (the metallic state), $\Delta\rho$ is the difference between 
the ionicity $\rho$ of the two dimers in the CO state,
and $\varDelta_i$ is the ionization frequency shift. 

Since we are considering the ($bc$) spectrum, we have to take into account that the $\Delta\rho$ dependence of $\nu_2$ and $\nu_3$ is not linear, being affected
by the inter-dimer e-mv coupling discussed in Section \ref{s:trimermodel}.
Accordingly, the e-mv perturbed frequencies $\Omega_i^\pm$ are obtained
from the diagonalization of the following  ``force constants'' matrix
\begin{equation}
	F_{ij}^\pm = \omega_i^\pm \omega_j^\pm \delta_{ij} - g_i g_j\chi_R(0) \sqrt{\omega_i^\pm \omega_j^\pm} 
	\label{eq:force_const}
\end{equation} 
where $\delta_{ij}$ is the Kronecker delta, $\omega_{i,j}^\pm$ and $g_{i,j}$ are the unperturbed frequencies and corresponding e-mv coupling constants \cite{Girlando2011}, 
and $\chi_R(0)$ the Raman susceptibility.

In the simulation we have added to the three C=C stretching modes
just one
CH bending in representation of the ones occurring in the spectral region
of interest \cite{Girlando2011}. The frequencies and parameters
used in the simulation are reported in Table B.1.

\begin{table}
	\label{tab:param}
\begin{center}
\caption{Parameters adopted in the phenomenological simulation}
\begin{tabular}{|l|c|c|c|}
	\hline
 & $\omega_i (0.5)$ (cm$^{-1}$) & $\varDelta_i$ (cm$^{-1}$) & $g_i$ (meV)	\\
 \hline
$a_g \nu_2$ & 1492.0 & 112.0 & 30.0 \\
$a_g \nu_3$ & 1479.0 & 120.0 & 71.0 \\
$b_{1u} \nu_{27}$ & 1468.0 & 140.0 & \ 0.0   \\
CH bend  &1447.0 &  \ \ \ 0.0 &  \ 0.0   \\
	\hline
\end{tabular} 
\end{center}
\end{table}
The bottom panel of Fig. \ref{fig:S_RamanCO} shows the results, plotted
as the $\Delta\rho$ dependence of the seven $\Omega_i^\pm$ frequencies.

All the modes included in the bottom panel of Fig. \ref{fig:S_RamanCO}
have the same symmetry. Therefore the lines expressing
their $\Delta \rho$ dependence must obey the non-crossing rule: They approach,
but when they become almost degenerate, by perturbation theory they mix and
repel each other. Consider first the highest frequency
modes, i.e., the red lines corresponding the $a_g \nu_2$ of the charge rich ($\nu_2^+$)
and charge poor ($\nu_2^-$) ET molecule. In the $\Delta \rho$
interval considered in the Figure, 
the frequencies have a practically linear dependence \cite{Yamamoto2005}. Comparison
with the upper panel of Fig. \ref{fig:S_RamanCO} immediately yields
the assignment of the $\nu_2$ mode to the band at 1504 \cm and to the shoulder at 1486 \cm, as done in Ref. \cite{Hassan2020}. The estimated $\Delta \rho = 0.2$, evidenced by the horizontal dotted line, is in agreement with the IR data \cite{Drichko2014}. Going towards lower frequencies, we have an avoided crossing and quasi-degeneracy between $\nu_{27}^+$ (blue) and $\nu_3^+$ (magenta), and both are assigned to the band seen at 1474 \cm (we have used two Lorentzians in the deconvolution of the top panel, although one would give an equally good fitting). Proceeding towards lower frequencies
we encounter the $\nu_{27}^-$, the CH bending (maroon line) and finally the e-mv
perturbed $\nu_3^-$. Therefore the adopted phenomenological approach gives
a satisfactory assignment of the C=C stretching phonons in the CO state
of \kHg.

\bibliography{/home/hal/Work/Articoli/CTcrystals.bib}

\begin{thebibliography}{35}%
\makeatletter
\providecommand \@ifxundefined [1]{%
 \@ifx{#1\undefined}
}%
\providecommand \@ifnum [1]{%
 \ifnum #1\expandafter \@firstoftwo
 \else \expandafter \@secondoftwo
 \fi
}%
\providecommand \@ifx [1]{%
 \ifx #1\expandafter \@firstoftwo
 \else \expandafter \@secondoftwo
 \fi
}%
\providecommand \natexlab [1]{#1}%
\providecommand \enquote  [1]{``#1''}%
\providecommand \bibnamefont  [1]{#1}%
\providecommand \bibfnamefont [1]{#1}%
\providecommand \citenamefont [1]{#1}%
\providecommand \href@noop [0]{\@secondoftwo}%
\providecommand \href [0]{\begingroup \@sanitize@url \@href}%
\providecommand \@href[1]{\@@startlink{#1}\@@href}%
\providecommand \@@href[1]{\endgroup#1\@@endlink}%
\providecommand \@sanitize@url [0]{\catcode `\\12\catcode `\$12\catcode
  `\&12\catcode `\#12\catcode `\^12\catcode `\_12\catcode `\%12\relax}%
\providecommand \@@startlink[1]{}%
\providecommand \@@endlink[0]{}%
\providecommand \url  [0]{\begingroup\@sanitize@url \@url }%
\providecommand \@url [1]{\endgroup\@href {#1}{\urlprefix }}%
\providecommand \urlprefix  [0]{URL }%
\providecommand \Eprint [0]{\href }%
\providecommand \doibase [0]{https://doi.org/}%
\providecommand \selectlanguage [0]{\@gobble}%
\providecommand \bibinfo  [0]{\@secondoftwo}%
\providecommand \bibfield  [0]{\@secondoftwo}%
\providecommand \translation [1]{[#1]}%
\providecommand \BibitemOpen [0]{}%
\providecommand \bibitemStop [0]{}%
\providecommand \bibitemNoStop [0]{.\EOS\space}%
\providecommand \EOS [0]{\spacefactor3000\relax}%
\providecommand \BibitemShut  [1]{\csname bibitem#1\endcsname}%
\let\auto@bib@innerbib\@empty
\bibitem [{\citenamefont {Powell}\ and\ \citenamefont
  {McKenzie}(2011)}]{Powell2011}%
  \BibitemOpen
  \bibfield  {author} {\bibinfo {author} {\bibfnamefont {B.~J.}\ \bibnamefont
  {Powell}}\ and\ \bibinfo {author} {\bibfnamefont {R.~H.}\ \bibnamefont
  {McKenzie}},\ }\href {https://doi.org/10.1088/0034-4885/74/5/056501}
  {\bibfield  {journal} {\bibinfo  {journal} {Reports on Progress in Physics}\
  }\textbf {\bibinfo {volume} {74}},\ \bibinfo {pages} {056501} (\bibinfo
  {year} {2011})}\BibitemShut {NoStop}%
\bibitem [{\citenamefont {Ardavan}\ \emph {et~al.}(2012)\citenamefont
  {Ardavan}, \citenamefont {Brown}, \citenamefont {Kagoshima}, \citenamefont
  {Kanoda}, \citenamefont {Kuroki}, \citenamefont {Mori}, \citenamefont
  {Ogata}, \citenamefont {Uji},\ and\ \citenamefont {Wosnitza}}]{Ardavan2012}%
  \BibitemOpen
  \bibfield  {author} {\bibinfo {author} {\bibfnamefont {A.}~\bibnamefont
  {Ardavan}}, \bibinfo {author} {\bibfnamefont {S.}~\bibnamefont {Brown}},
  \bibinfo {author} {\bibfnamefont {S.}~\bibnamefont {Kagoshima}}, \bibinfo
  {author} {\bibfnamefont {K.}~\bibnamefont {Kanoda}}, \bibinfo {author}
  {\bibfnamefont {K.}~\bibnamefont {Kuroki}}, \bibinfo {author} {\bibfnamefont
  {H.}~\bibnamefont {Mori}}, \bibinfo {author} {\bibfnamefont {M.}~\bibnamefont
  {Ogata}}, \bibinfo {author} {\bibfnamefont {S.}~\bibnamefont {Uji}},\ and\
  \bibinfo {author} {\bibfnamefont {J.}~\bibnamefont {Wosnitza}},\ }\href
  {https://doi.org/10.1143/jpsj.81.011004} {\bibfield  {journal} {\bibinfo
  {journal} {Journal of the Physical Society of Japan}\ }\textbf {\bibinfo
  {volume} {81}},\ \bibinfo {pages} {011004} (\bibinfo {year}
  {2012})}\BibitemShut {NoStop}%
\bibitem [{\citenamefont {Wosnitza}(2012)}]{Wosnitza2012}%
  \BibitemOpen
  \bibfield  {author} {\bibinfo {author} {\bibfnamefont {J.}~\bibnamefont
  {Wosnitza}},\ }\href {https://doi.org/10.3390/cryst2020248} {\bibfield
  {journal} {\bibinfo  {journal} {Crystals}\ }\textbf {\bibinfo {volume} {2}},\
  \bibinfo {pages} {248} (\bibinfo {year} {2012})}\BibitemShut {NoStop}%
\bibitem [{\citenamefont {Dressel}\ and\ \citenamefont
  {Tomi{\'{c}}}(2020)}]{Dressel2020}%
  \BibitemOpen
  \bibfield  {author} {\bibinfo {author} {\bibfnamefont {M.}~\bibnamefont
  {Dressel}}\ and\ \bibinfo {author} {\bibfnamefont {S.}~\bibnamefont
  {Tomi{\'{c}}}},\ }\href {https://doi.org/10.1080/00018732.2020.1837833}
  {\bibfield  {journal} {\bibinfo  {journal} {Advances in Physics}\ }\textbf
  {\bibinfo {volume} {69}},\ \bibinfo {pages} {1} (\bibinfo {year}
  {2020})}\BibitemShut {NoStop}%
\bibitem [{\citenamefont {Riedl}\ \emph {et~al.}(2022)\citenamefont {Riedl},
  \citenamefont {Gati},\ and\ \citenamefont {Valentí}}]{Riedl2022}%
  \BibitemOpen
  \bibfield  {author} {\bibinfo {author} {\bibfnamefont {K.}~\bibnamefont
  {Riedl}}, \bibinfo {author} {\bibfnamefont {E.}~\bibnamefont {Gati}},\ and\
  \bibinfo {author} {\bibfnamefont {R.}~\bibnamefont {Valentí}},\ }\href
  {https://doi.org/10.3390/cryst12121689} {\bibfield  {journal} {\bibinfo
  {journal} {Crystals}\ }\textbf {\bibinfo {volume} {12}},\ \bibinfo {pages}
  {1689} (\bibinfo {year} {2022})}\BibitemShut {NoStop}%
\bibitem [{\citenamefont {Mori}\ \emph {et~al.}(1984)\citenamefont {Mori},
  \citenamefont {Kobayashi}, \citenamefont {Sasaki}, \citenamefont {Kobayashi},
  \citenamefont {Saito},\ and\ \citenamefont {Inokuchi}}]{Mori1984}%
  \BibitemOpen
  \bibfield  {author} {\bibinfo {author} {\bibfnamefont {T.}~\bibnamefont
  {Mori}}, \bibinfo {author} {\bibfnamefont {A.}~\bibnamefont {Kobayashi}},
  \bibinfo {author} {\bibfnamefont {Y.}~\bibnamefont {Sasaki}}, \bibinfo
  {author} {\bibfnamefont {H.}~\bibnamefont {Kobayashi}}, \bibinfo {author}
  {\bibfnamefont {G.}~\bibnamefont {Saito}},\ and\ \bibinfo {author}
  {\bibfnamefont {H.}~\bibnamefont {Inokuchi}},\ }\href
  {https://doi.org/10.1246/bcsj.57.627} {\bibfield  {journal} {\bibinfo
  {journal} {Bulletin of the Chemical Society of Japan}\ }\textbf {\bibinfo
  {volume} {57}},\ \bibinfo {pages} {627} (\bibinfo {year} {1984})}\BibitemShut
  {NoStop}%
\bibitem [{\citenamefont {Mori}(1998)}]{Mori1998}%
  \BibitemOpen
  \bibfield  {author} {\bibinfo {author} {\bibfnamefont {T.}~\bibnamefont
  {Mori}},\ }\href {https://doi.org/10.1246/bcsj.71.2509} {\bibfield  {journal}
  {\bibinfo  {journal} {Bulletin of the Chemical Society of Japan}\ }\textbf
  {\bibinfo {volume} {71}},\ \bibinfo {pages} {2509} (\bibinfo {year}
  {1998})}\BibitemShut {NoStop}%
\bibitem [{\citenamefont {Mori}\ \emph {et~al.}(1999)\citenamefont {Mori},
  \citenamefont {Mori},\ and\ \citenamefont {Tanaka}}]{Mori1999}%
  \BibitemOpen
  \bibfield  {author} {\bibinfo {author} {\bibfnamefont {T.}~\bibnamefont
  {Mori}}, \bibinfo {author} {\bibfnamefont {H.}~\bibnamefont {Mori}},\ and\
  \bibinfo {author} {\bibfnamefont {S.}~\bibnamefont {Tanaka}},\ }\href
  {https://doi.org/10.1246/bcsj.72.179} {\bibfield  {journal} {\bibinfo
  {journal} {Bulletin of the Chemical Society of Japan}\ }\textbf {\bibinfo
  {volume} {72}},\ \bibinfo {pages} {179} (\bibinfo {year} {1999})}\BibitemShut
  {NoStop}%
\bibitem [{\citenamefont {Yamamoto}\ \emph {et~al.}(2005)\citenamefont
  {Yamamoto}, \citenamefont {Uruichi}, \citenamefont {Yamamoto}, \citenamefont
  {Yakushi}, \citenamefont {Kawamoto},\ and\ \citenamefont
  {Taniguchi}}]{Yamamoto2005}%
  \BibitemOpen
  \bibfield  {author} {\bibinfo {author} {\bibfnamefont {T.}~\bibnamefont
  {Yamamoto}}, \bibinfo {author} {\bibfnamefont {M.}~\bibnamefont {Uruichi}},
  \bibinfo {author} {\bibfnamefont {K.}~\bibnamefont {Yamamoto}}, \bibinfo
  {author} {\bibfnamefont {K.}~\bibnamefont {Yakushi}}, \bibinfo {author}
  {\bibfnamefont {A.}~\bibnamefont {Kawamoto}},\ and\ \bibinfo {author}
  {\bibfnamefont {H.}~\bibnamefont {Taniguchi}},\ }\href
  {https://doi.org/10.1021/jp050247o} {\bibfield  {journal} {\bibinfo
  {journal} {The Journal of Physical Chemistry B}\ }\textbf {\bibinfo {volume}
  {109}},\ \bibinfo {pages} {15226} (\bibinfo {year} {2005})}\BibitemShut
  {NoStop}%
\bibitem [{\citenamefont {Girlando}(2011)}]{Girlando2011}%
  \BibitemOpen
  \bibfield  {author} {\bibinfo {author} {\bibfnamefont {A.}~\bibnamefont
  {Girlando}},\ }\href {https://doi.org/10.1021/jp206171r} {\bibfield
  {journal} {\bibinfo  {journal} {Journal of Physical Chemistry C}\ }\textbf
  {\bibinfo {volume} {115}},\ \bibinfo {pages} {19371} (\bibinfo {year}
  {2011})}\BibitemShut {NoStop}%
\bibitem [{\citenamefont {Rice}(1979)}]{Rice1979}%
  \BibitemOpen
  \bibfield  {author} {\bibinfo {author} {\bibfnamefont {M.}~\bibnamefont
  {Rice}},\ }\href {https://doi.org/10.1016/0038-1098(79)90175-3} {\bibfield
  {journal} {\bibinfo  {journal} {Solid State Communications}\ }\textbf
  {\bibinfo {volume} {31}},\ \bibinfo {pages} {93} (\bibinfo {year}
  {1979})}\BibitemShut {NoStop}%
\bibitem [{\citenamefont {Painelli}\ and\ \citenamefont
  {Girlando}(1986)}]{Painelli1986}%
  \BibitemOpen
  \bibfield  {author} {\bibinfo {author} {\bibfnamefont {A.}~\bibnamefont
  {Painelli}}\ and\ \bibinfo {author} {\bibfnamefont {A.}~\bibnamefont
  {Girlando}},\ }\href {https://doi.org/10.1063/1.449926} {\bibfield  {journal}
  {\bibinfo  {journal} {The Journal of Chemical Physics}\ }\textbf {\bibinfo
  {volume} {84}},\ \bibinfo {pages} {5655} (\bibinfo {year}
  {1986})}\BibitemShut {NoStop}%
\bibitem [{\citenamefont {Girlando}\ \emph {et~al.}(2012)\citenamefont
  {Girlando}, \citenamefont {Masino}, \citenamefont {Kaiser}, \citenamefont
  {Sun}, \citenamefont {Drichko}, \citenamefont {Dressel},\ and\ \citenamefont
  {Mori}}]{Girlando2012}%
  \BibitemOpen
  \bibfield  {author} {\bibinfo {author} {\bibfnamefont {A.}~\bibnamefont
  {Girlando}}, \bibinfo {author} {\bibfnamefont {M.}~\bibnamefont {Masino}},
  \bibinfo {author} {\bibfnamefont {S.}~\bibnamefont {Kaiser}}, \bibinfo
  {author} {\bibfnamefont {Y.}~\bibnamefont {Sun}}, \bibinfo {author}
  {\bibfnamefont {N.}~\bibnamefont {Drichko}}, \bibinfo {author} {\bibfnamefont
  {M.}~\bibnamefont {Dressel}},\ and\ \bibinfo {author} {\bibfnamefont
  {H.}~\bibnamefont {Mori}},\ }\href {https://doi.org/10.1002/pssb.201100722}
  {\bibfield  {journal} {\bibinfo  {journal} {Physica Status Solidi (B) Basic
  Research}\ }\textbf {\bibinfo {volume} {249}},\ \bibinfo {pages} {953}
  (\bibinfo {year} {2012})}\BibitemShut {NoStop}%
\bibitem [{\citenamefont {Girlando}\ \emph {et~al.}(2014)\citenamefont
  {Girlando}, \citenamefont {Masino}, \citenamefont {Schlueter}, \citenamefont
  {Drichko}, \citenamefont {Kaiser},\ and\ \citenamefont
  {Dressel}}]{Girlando2014}%
  \BibitemOpen
  \bibfield  {author} {\bibinfo {author} {\bibfnamefont {A.}~\bibnamefont
  {Girlando}}, \bibinfo {author} {\bibfnamefont {M.}~\bibnamefont {Masino}},
  \bibinfo {author} {\bibfnamefont {J.~A.}\ \bibnamefont {Schlueter}}, \bibinfo
  {author} {\bibfnamefont {N.}~\bibnamefont {Drichko}}, \bibinfo {author}
  {\bibfnamefont {S.}~\bibnamefont {Kaiser}},\ and\ \bibinfo {author}
  {\bibfnamefont {M.}~\bibnamefont {Dressel}},\ }\href
  {https://doi.org/10.1103/physrevb.89.174503} {\bibfield  {journal} {\bibinfo
  {journal} {Physical Review B}\ }\textbf {\bibinfo {volume} {89}},\ \bibinfo
  {pages} {174503} (\bibinfo {year} {2014})}\BibitemShut {NoStop}%
\bibitem [{\citenamefont {Maksimuk}\ \emph {et~al.}(2001)\citenamefont
  {Maksimuk}, \citenamefont {Yakushi}, \citenamefont {Taniguchi}, \citenamefont
  {Kanoda},\ and\ \citenamefont {Kawamoto}}]{Maksimuk2001}%
  \BibitemOpen
  \bibfield  {author} {\bibinfo {author} {\bibfnamefont {M.}~\bibnamefont
  {Maksimuk}}, \bibinfo {author} {\bibfnamefont {K.}~\bibnamefont {Yakushi}},
  \bibinfo {author} {\bibfnamefont {H.}~\bibnamefont {Taniguchi}}, \bibinfo
  {author} {\bibfnamefont {K.}~\bibnamefont {Kanoda}},\ and\ \bibinfo {author}
  {\bibfnamefont {A.}~\bibnamefont {Kawamoto}},\ }\href
  {https://doi.org/10.1143/jpsj.70.3728} {\bibfield  {journal} {\bibinfo
  {journal} {Journal of the Physical Society of Japan}\ }\textbf {\bibinfo
  {volume} {70}},\ \bibinfo {pages} {3728} (\bibinfo {year}
  {2001})}\BibitemShut {NoStop}%
\bibitem [{\citenamefont {Yamamoto}\ \emph {et~al.}(2021)\citenamefont
  {Yamamoto}, \citenamefont {Naito}, \citenamefont {Uruichi}, \citenamefont
  {Akutsu},\ and\ \citenamefont {Nakazawa}}]{Yamamoto2021}%
  \BibitemOpen
  \bibfield  {author} {\bibinfo {author} {\bibfnamefont {T.}~\bibnamefont
  {Yamamoto}}, \bibinfo {author} {\bibfnamefont {T.}~\bibnamefont {Naito}},
  \bibinfo {author} {\bibfnamefont {M.}~\bibnamefont {Uruichi}}, \bibinfo
  {author} {\bibfnamefont {H.}~\bibnamefont {Akutsu}},\ and\ \bibinfo {author}
  {\bibfnamefont {Y.}~\bibnamefont {Nakazawa}},\ }\href
  {https://doi.org/10.7566/jpsj.90.063708} {\bibfield  {journal} {\bibinfo
  {journal} {Journal of the Physical Society of Japan}\ }\textbf {\bibinfo
  {volume} {90}},\ \bibinfo {pages} {063708} (\bibinfo {year}
  {2021})}\BibitemShut {NoStop}%
\bibitem [{\citenamefont {Hassan}\ \emph {et~al.}(2020)\citenamefont {Hassan},
  \citenamefont {Thirunavukkuarasu}, \citenamefont {Lu}, \citenamefont
  {Smirnov}, \citenamefont {Zhilyaeva}, \citenamefont {Torunova}, \citenamefont
  {Lyubovskaya},\ and\ \citenamefont {Drichko}}]{Hassan2020}%
  \BibitemOpen
  \bibfield  {author} {\bibinfo {author} {\bibfnamefont {N.~M.}\ \bibnamefont
  {Hassan}}, \bibinfo {author} {\bibfnamefont {K.}~\bibnamefont
  {Thirunavukkuarasu}}, \bibinfo {author} {\bibfnamefont {Z.}~\bibnamefont
  {Lu}}, \bibinfo {author} {\bibfnamefont {D.}~\bibnamefont {Smirnov}},
  \bibinfo {author} {\bibfnamefont {E.~I.}\ \bibnamefont {Zhilyaeva}}, \bibinfo
  {author} {\bibfnamefont {S.}~\bibnamefont {Torunova}}, \bibinfo {author}
  {\bibfnamefont {R.~N.}\ \bibnamefont {Lyubovskaya}},\ and\ \bibinfo {author}
  {\bibfnamefont {N.}~\bibnamefont {Drichko}},\ }\href
  {https://doi.org/10.1038/s41535-020-0217-5} {\bibfield  {journal} {\bibinfo
  {journal} {npj Quantum Materials}\ }\textbf {\bibinfo {volume} {5}},\
  \bibinfo {pages} {15} (\bibinfo {year} {2020})}\BibitemShut {NoStop}%
\bibitem [{\citenamefont {Drichko}\ \emph {et~al.}(2014)\citenamefont
  {Drichko}, \citenamefont {Beyer}, \citenamefont {Rose}, \citenamefont
  {Dressel}, \citenamefont {Schlueter}, \citenamefont {Turunova}, \citenamefont
  {Zhilyaeva},\ and\ \citenamefont {Lyubovskaya}}]{Drichko2014}%
  \BibitemOpen
  \bibfield  {author} {\bibinfo {author} {\bibfnamefont {N.}~\bibnamefont
  {Drichko}}, \bibinfo {author} {\bibfnamefont {R.}~\bibnamefont {Beyer}},
  \bibinfo {author} {\bibfnamefont {E.}~\bibnamefont {Rose}}, \bibinfo {author}
  {\bibfnamefont {M.}~\bibnamefont {Dressel}}, \bibinfo {author} {\bibfnamefont
  {J.~A.}\ \bibnamefont {Schlueter}}, \bibinfo {author} {\bibfnamefont {S.~A.}\
  \bibnamefont {Turunova}}, \bibinfo {author} {\bibfnamefont {E.~I.}\
  \bibnamefont {Zhilyaeva}},\ and\ \bibinfo {author} {\bibfnamefont {R.~N.}\
  \bibnamefont {Lyubovskaya}},\ }\href
  {https://doi.org/10.1103/physrevb.89.075133} {\bibfield  {journal} {\bibinfo
  {journal} {Physical Review B}\ }\textbf {\bibinfo {volume} {89}},\ \bibinfo
  {pages} {075133} (\bibinfo {year} {2014})}\BibitemShut {NoStop}%
\bibitem [{\citenamefont {Löhle}\ \emph {et~al.}(2016)\citenamefont {Löhle},
  \citenamefont {Rose}, \citenamefont {Singh}, \citenamefont {Beyer},
  \citenamefont {Tafra}, \citenamefont {Ivek}, \citenamefont {Zhilyaeva},
  \citenamefont {Lyubovskaya},\ and\ \citenamefont {Dressel}}]{Loehle2016}%
  \BibitemOpen
  \bibfield  {author} {\bibinfo {author} {\bibfnamefont {A.}~\bibnamefont
  {Löhle}}, \bibinfo {author} {\bibfnamefont {E.}~\bibnamefont {Rose}},
  \bibinfo {author} {\bibfnamefont {S.}~\bibnamefont {Singh}}, \bibinfo
  {author} {\bibfnamefont {R.}~\bibnamefont {Beyer}}, \bibinfo {author}
  {\bibfnamefont {E.}~\bibnamefont {Tafra}}, \bibinfo {author} {\bibfnamefont
  {T.}~\bibnamefont {Ivek}}, \bibinfo {author} {\bibfnamefont {E.~I.}\
  \bibnamefont {Zhilyaeva}}, \bibinfo {author} {\bibfnamefont {R.~N.}\
  \bibnamefont {Lyubovskaya}},\ and\ \bibinfo {author} {\bibfnamefont
  {M.}~\bibnamefont {Dressel}},\ }\href
  {https://doi.org/10.1088/1361-648x/29/5/055601} {\bibfield  {journal}
  {\bibinfo  {journal} {Journal of Physics: Condensed Matter}\ }\textbf
  {\bibinfo {volume} {29}},\ \bibinfo {pages} {055601} (\bibinfo {year}
  {2016})}\BibitemShut {NoStop}%
\bibitem [{\citenamefont {Gati}\ \emph {et~al.}(2018)\citenamefont {Gati},
  \citenamefont {Fischer}, \citenamefont {Lunkenheimer}, \citenamefont
  {Zielke}, \citenamefont {Köhler}, \citenamefont {Kolb}, \citenamefont {von
  Nidda}, \citenamefont {Winter}, \citenamefont {Schubert}, \citenamefont
  {Schlueter}, \citenamefont {Jeschke}, \citenamefont {Valent{\'{\i}}},\ and\
  \citenamefont {Lang}}]{Gati2018}%
  \BibitemOpen
  \bibfield  {author} {\bibinfo {author} {\bibfnamefont {E.}~\bibnamefont
  {Gati}}, \bibinfo {author} {\bibfnamefont {J.~K.~H.}\ \bibnamefont
  {Fischer}}, \bibinfo {author} {\bibfnamefont {P.}~\bibnamefont
  {Lunkenheimer}}, \bibinfo {author} {\bibfnamefont {D.}~\bibnamefont
  {Zielke}}, \bibinfo {author} {\bibfnamefont {S.}~\bibnamefont {Köhler}},
  \bibinfo {author} {\bibfnamefont {F.}~\bibnamefont {Kolb}}, \bibinfo {author}
  {\bibfnamefont {H.-A.~K.}\ \bibnamefont {von Nidda}}, \bibinfo {author}
  {\bibfnamefont {S.~M.}\ \bibnamefont {Winter}}, \bibinfo {author}
  {\bibfnamefont {H.}~\bibnamefont {Schubert}}, \bibinfo {author}
  {\bibfnamefont {J.~A.}\ \bibnamefont {Schlueter}}, \bibinfo {author}
  {\bibfnamefont {H.~O.}\ \bibnamefont {Jeschke}}, \bibinfo {author}
  {\bibfnamefont {R.}~\bibnamefont {Valent{\'{\i}}}},\ and\ \bibinfo {author}
  {\bibfnamefont {M.}~\bibnamefont {Lang}},\ }\href
  {https://doi.org/10.1103/physrevlett.120.247601} {\bibfield  {journal}
  {\bibinfo  {journal} {Physical Review Letters}\ }\textbf {\bibinfo {volume}
  {120}},\ \bibinfo {pages} {247601} (\bibinfo {year} {2018})}\BibitemShut
  {NoStop}%
\bibitem [{\citenamefont {Turrell}(1972)}]{Turrell1972}%
  \BibitemOpen
  \bibfield  {author} {\bibinfo {author} {\bibfnamefont {G.}~\bibnamefont
  {Turrell}},\ }\href@noop {} {\emph {\bibinfo {title} {\textit{Infrared and
  Raman Spectra of Crystals}}}}\ (\bibinfo  {publisher} {Academic Press},\
  \bibinfo {address} {London and New York},\ \bibinfo {year}
  {1972})\BibitemShut {NoStop}%
\bibitem [{\citenamefont {Barca}\ \emph {et~al.}(2020)\citenamefont {Barca},
  \citenamefont {Bertoni}, \citenamefont {Carrington}, \citenamefont {Datta},
  \citenamefont {Silva}, \citenamefont {Deustua}, \citenamefont {Fedorov},
  \citenamefont {Gour}, \citenamefont {Gunina}, \citenamefont {Guidez},
  \citenamefont {Harville}, \citenamefont {Irle}, \citenamefont {Ivanic},
  \citenamefont {Kowalski}, \citenamefont {Leang}, \citenamefont {Li},
  \citenamefont {Li}, \citenamefont {Lutz}, \citenamefont {Magoulas},
  \citenamefont {Mato}, \citenamefont {Mironov}, \citenamefont {Nakata},
  \citenamefont {Pham}, \citenamefont {Piecuch}, \citenamefont {Poole},
  \citenamefont {Pruitt}, \citenamefont {Rendell}, \citenamefont {Roskop},
  \citenamefont {Ruedenberg}, \citenamefont {Sattasathuchana}, \citenamefont
  {Schmidt}, \citenamefont {Shen}, \citenamefont {Slipchenko}, \citenamefont
  {Sosonkina}, \citenamefont {Sundriyal}, \citenamefont {Tiwari}, \citenamefont
  {Vallejo}, \citenamefont {Westheimer}, \citenamefont {W{\l}och},
  \citenamefont {Xu}, \citenamefont {Zahariev},\ and\ \citenamefont
  {Gordon}}]{GAMESS}%
  \BibitemOpen
  \bibfield  {author} {\bibinfo {author} {\bibfnamefont {G.~M.~J.}\
  \bibnamefont {Barca}}, \bibinfo {author} {\bibfnamefont {C.}~\bibnamefont
  {Bertoni}}, \bibinfo {author} {\bibfnamefont {L.}~\bibnamefont {Carrington}},
  \bibinfo {author} {\bibfnamefont {D.}~\bibnamefont {Datta}}, \bibinfo
  {author} {\bibfnamefont {N.~D.}\ \bibnamefont {Silva}}, \bibinfo {author}
  {\bibfnamefont {J.~E.}\ \bibnamefont {Deustua}}, \bibinfo {author}
  {\bibfnamefont {D.~G.}\ \bibnamefont {Fedorov}}, \bibinfo {author}
  {\bibfnamefont {J.~R.}\ \bibnamefont {Gour}}, \bibinfo {author}
  {\bibfnamefont {A.~O.}\ \bibnamefont {Gunina}}, \bibinfo {author}
  {\bibfnamefont {E.}~\bibnamefont {Guidez}}, \bibinfo {author} {\bibfnamefont
  {T.}~\bibnamefont {Harville}}, \bibinfo {author} {\bibfnamefont
  {S.}~\bibnamefont {Irle}}, \bibinfo {author} {\bibfnamefont {J.}~\bibnamefont
  {Ivanic}}, \bibinfo {author} {\bibfnamefont {K.}~\bibnamefont {Kowalski}},
  \bibinfo {author} {\bibfnamefont {S.~S.}\ \bibnamefont {Leang}}, \bibinfo
  {author} {\bibfnamefont {H.}~\bibnamefont {Li}}, \bibinfo {author}
  {\bibfnamefont {W.}~\bibnamefont {Li}}, \bibinfo {author} {\bibfnamefont
  {J.~J.}\ \bibnamefont {Lutz}}, \bibinfo {author} {\bibfnamefont
  {I.}~\bibnamefont {Magoulas}}, \bibinfo {author} {\bibfnamefont
  {J.}~\bibnamefont {Mato}}, \bibinfo {author} {\bibfnamefont {V.}~\bibnamefont
  {Mironov}}, \bibinfo {author} {\bibfnamefont {H.}~\bibnamefont {Nakata}},
  \bibinfo {author} {\bibfnamefont {B.~Q.}\ \bibnamefont {Pham}}, \bibinfo
  {author} {\bibfnamefont {P.}~\bibnamefont {Piecuch}}, \bibinfo {author}
  {\bibfnamefont {D.}~\bibnamefont {Poole}}, \bibinfo {author} {\bibfnamefont
  {S.~R.}\ \bibnamefont {Pruitt}}, \bibinfo {author} {\bibfnamefont {A.~P.}\
  \bibnamefont {Rendell}}, \bibinfo {author} {\bibfnamefont {L.~B.}\
  \bibnamefont {Roskop}}, \bibinfo {author} {\bibfnamefont {K.}~\bibnamefont
  {Ruedenberg}}, \bibinfo {author} {\bibfnamefont {T.}~\bibnamefont
  {Sattasathuchana}}, \bibinfo {author} {\bibfnamefont {M.~W.}\ \bibnamefont
  {Schmidt}}, \bibinfo {author} {\bibfnamefont {J.}~\bibnamefont {Shen}},
  \bibinfo {author} {\bibfnamefont {L.}~\bibnamefont {Slipchenko}}, \bibinfo
  {author} {\bibfnamefont {M.}~\bibnamefont {Sosonkina}}, \bibinfo {author}
  {\bibfnamefont {V.}~\bibnamefont {Sundriyal}}, \bibinfo {author}
  {\bibfnamefont {A.}~\bibnamefont {Tiwari}}, \bibinfo {author} {\bibfnamefont
  {J.~L.~G.}\ \bibnamefont {Vallejo}}, \bibinfo {author} {\bibfnamefont
  {B.}~\bibnamefont {Westheimer}}, \bibinfo {author} {\bibfnamefont
  {M.}~\bibnamefont {W{\l}och}}, \bibinfo {author} {\bibfnamefont
  {P.}~\bibnamefont {Xu}}, \bibinfo {author} {\bibfnamefont {F.}~\bibnamefont
  {Zahariev}},\ and\ \bibinfo {author} {\bibfnamefont {M.~S.}\ \bibnamefont
  {Gordon}},\ }\href {https://doi.org/10.1063/5.0005188} {\bibfield  {journal}
  {\bibinfo  {journal} {The Journal of Chemical Physics}\ }\textbf {\bibinfo
  {volume} {152}},\ \bibinfo {pages} {154102} (\bibinfo {year}
  {2020})}\BibitemShut {NoStop}%
\bibitem [{\citenamefont {Yakushi}\ \emph {et~al.}(2015)\citenamefont
  {Yakushi}, \citenamefont {Yamamoto}, \citenamefont {Yamamoto}, \citenamefont
  {Saito},\ and\ \citenamefont {Kawamoto}}]{Yakushi2015}%
  \BibitemOpen
  \bibfield  {author} {\bibinfo {author} {\bibfnamefont {K.}~\bibnamefont
  {Yakushi}}, \bibinfo {author} {\bibfnamefont {K.}~\bibnamefont {Yamamoto}},
  \bibinfo {author} {\bibfnamefont {T.}~\bibnamefont {Yamamoto}}, \bibinfo
  {author} {\bibfnamefont {Y.}~\bibnamefont {Saito}},\ and\ \bibinfo {author}
  {\bibfnamefont {A.}~\bibnamefont {Kawamoto}},\ }\href
  {https://doi.org/10.7566/jpsj.84.084711} {\bibfield  {journal} {\bibinfo
  {journal} {Journal of the Physical Society of Japan}\ }\textbf {\bibinfo
  {volume} {84}},\ \bibinfo {pages} {084711} (\bibinfo {year}
  {2015})}\BibitemShut {NoStop}%
\bibitem [{\citenamefont {Yartsev}(1982)}]{Yartsev1982}%
  \BibitemOpen
  \bibfield  {author} {\bibinfo {author} {\bibfnamefont {V.~M.}\ \bibnamefont
  {Yartsev}},\ }\href@noop {} {\bibfield  {journal} {\bibinfo  {journal} {Phys.
  Stat. Sol. (b) 112, 279 (1982)}\ }\textbf {\bibinfo {volume} {112}},\
  \bibinfo {pages} {279} (\bibinfo {year} {1982})}\BibitemShut {NoStop}%
\bibitem [{\citenamefont {Painelli}\ \emph {et~al.}(1986)\citenamefont
  {Painelli}, \citenamefont {Pecile},\ and\ \citenamefont
  {Girlando}}]{Painelli1986a}%
  \BibitemOpen
  \bibfield  {author} {\bibinfo {author} {\bibfnamefont {A.}~\bibnamefont
  {Painelli}}, \bibinfo {author} {\bibfnamefont {C.}~\bibnamefont {Pecile}},\
  and\ \bibinfo {author} {\bibfnamefont {A.}~\bibnamefont {Girlando}},\ }\href
  {https://doi.org/10.1080/00268948608079572} {\bibfield  {journal} {\bibinfo
  {journal} {Molecular Crystals and Liquid Crystals}\ }\textbf {\bibinfo
  {volume} {134}},\ \bibinfo {pages} {1} (\bibinfo {year} {1986})}\BibitemShut
  {NoStop}%
\bibitem [{\citenamefont {Pecile}\ \emph {et~al.}(1989)\citenamefont {Pecile},
  \citenamefont {Painelli},\ and\ \citenamefont {Girlando}}]{Pecile1989}%
  \BibitemOpen
  \bibfield  {author} {\bibinfo {author} {\bibfnamefont {C.}~\bibnamefont
  {Pecile}}, \bibinfo {author} {\bibfnamefont {A.}~\bibnamefont {Painelli}},\
  and\ \bibinfo {author} {\bibfnamefont {A.}~\bibnamefont {Girlando}},\ }\href
  {https://doi.org/10.1080/00268948908065787} {\bibfield  {journal} {\bibinfo
  {journal} {Molecular Crystals and Liquid Crystals Incorporating Nonlinear
  Optics}\ }\textbf {\bibinfo {volume} {171}},\ \bibinfo {pages} {69} (\bibinfo
  {year} {1989})}\BibitemShut {NoStop}%
\bibitem [{\citenamefont {Revelli~Beaumont}\ \emph {et~al.}(2021)\citenamefont
  {Revelli~Beaumont}, \citenamefont {Hemme}, \citenamefont {Gallais},
  \citenamefont {Sacuto}, \citenamefont {Jacob}, \citenamefont {Valade},
  \citenamefont {de~Caro}, \citenamefont {Faulmann},\ and\ \citenamefont
  {Cazayous}}]{Revelli2021}%
  \BibitemOpen
  \bibfield  {author} {\bibinfo {author} {\bibfnamefont {M.}~\bibnamefont
  {Revelli~Beaumont}}, \bibinfo {author} {\bibfnamefont {P.}~\bibnamefont
  {Hemme}}, \bibinfo {author} {\bibfnamefont {Y.}~\bibnamefont {Gallais}},
  \bibinfo {author} {\bibfnamefont {A.}~\bibnamefont {Sacuto}}, \bibinfo
  {author} {\bibfnamefont {K.}~\bibnamefont {Jacob}}, \bibinfo {author}
  {\bibfnamefont {L.}~\bibnamefont {Valade}}, \bibinfo {author} {\bibfnamefont
  {D.}~\bibnamefont {de~Caro}}, \bibinfo {author} {\bibfnamefont
  {C.}~\bibnamefont {Faulmann}},\ and\ \bibinfo {author} {\bibfnamefont
  {M.}~\bibnamefont {Cazayous}},\ }\href
  {https://doi.org/10.1088/1361-648x/abd813} {\bibfield  {journal} {\bibinfo
  {journal} {Journal of Physics: Condensed Matter}\ }\textbf {\bibinfo {volume}
  {33}},\ \bibinfo {pages} {125403} (\bibinfo {year} {2021})}\BibitemShut
  {NoStop}%
\bibitem [{\citenamefont {Tamura}\ \emph {et~al.}(1991)\citenamefont {Tamura},
  \citenamefont {Tajima}, \citenamefont {Yakushi}, \citenamefont {Kuroda},
  \citenamefont {Kobayashi}, \citenamefont {Kato},\ and\ \citenamefont
  {Kobayashi}}]{Tamura1991}%
  \BibitemOpen
  \bibfield  {author} {\bibinfo {author} {\bibfnamefont {M.}~\bibnamefont
  {Tamura}}, \bibinfo {author} {\bibfnamefont {H.}~\bibnamefont {Tajima}},
  \bibinfo {author} {\bibfnamefont {K.}~\bibnamefont {Yakushi}}, \bibinfo
  {author} {\bibfnamefont {H.}~\bibnamefont {Kuroda}}, \bibinfo {author}
  {\bibfnamefont {A.}~\bibnamefont {Kobayashi}}, \bibinfo {author}
  {\bibfnamefont {R.}~\bibnamefont {Kato}},\ and\ \bibinfo {author}
  {\bibfnamefont {H.}~\bibnamefont {Kobayashi}},\ }\href
  {https://doi.org/10.1143/JPSJ.60.3861} {\bibfield  {journal} {\bibinfo
  {journal} {Journal of the Physical Society of Japan}\ }\textbf {\bibinfo
  {volume} {60}},\ \bibinfo {pages} {3861} (\bibinfo {year}
  {1991})}\BibitemShut {NoStop}%
\bibitem [{\citenamefont {Kino}\ and\ \citenamefont
  {Fukuyama}(1995)}]{Kino1995}%
  \BibitemOpen
  \bibfield  {author} {\bibinfo {author} {\bibfnamefont {H.}~\bibnamefont
  {Kino}}\ and\ \bibinfo {author} {\bibfnamefont {H.}~\bibnamefont
  {Fukuyama}},\ }\href {https://doi.org/10.1143/JPSJ.64.2726} {\bibfield
  {journal} {\bibinfo  {journal} {Journal of the Physical Society of Japan}\
  }\textbf {\bibinfo {volume} {64}},\ \bibinfo {pages} {2726} (\bibinfo {year}
  {1995})}\BibitemShut {NoStop}%
\bibitem [{\citenamefont {McKenzie}(1999)}]{McKenzie1999}%
  \BibitemOpen
  \bibfield  {author} {\bibinfo {author} {\bibfnamefont {R.~H.}\ \bibnamefont
  {McKenzie}},\ }\Eprint {https://arxiv.org/abs/cond-mat/9802198}
  {arXiv:cond-mat/9802198}  (\bibinfo {year} {1999})\BibitemShut {NoStop}%
\bibitem [{\citenamefont {Sedlmeier}\ \emph {et~al.}(2012)\citenamefont
  {Sedlmeier}, \citenamefont {Elsässer}, \citenamefont {Neubauer},
  \citenamefont {Beyer}, \citenamefont {Wu}, \citenamefont {Ivek},
  \citenamefont {Tomi{\'{c}}}, \citenamefont {Schlueter},\ and\ \citenamefont
  {Dressel}}]{Sedlmeier2012}%
  \BibitemOpen
  \bibfield  {author} {\bibinfo {author} {\bibfnamefont {K.}~\bibnamefont
  {Sedlmeier}}, \bibinfo {author} {\bibfnamefont {S.}~\bibnamefont
  {Elsässer}}, \bibinfo {author} {\bibfnamefont {D.}~\bibnamefont {Neubauer}},
  \bibinfo {author} {\bibfnamefont {R.}~\bibnamefont {Beyer}}, \bibinfo
  {author} {\bibfnamefont {D.}~\bibnamefont {Wu}}, \bibinfo {author}
  {\bibfnamefont {T.}~\bibnamefont {Ivek}}, \bibinfo {author} {\bibfnamefont
  {S.}~\bibnamefont {Tomi{\'{c}}}}, \bibinfo {author} {\bibfnamefont {J.~A.}\
  \bibnamefont {Schlueter}},\ and\ \bibinfo {author} {\bibfnamefont
  {M.}~\bibnamefont {Dressel}},\ }\href
  {https://doi.org/10.1103/PhysRevB.86.245103} {\bibfield  {journal} {\bibinfo
  {journal} {Physical Review B}\ }\textbf {\bibinfo {volume} {86}},\ \bibinfo
  {pages} {245103} (\bibinfo {year} {2012})}\BibitemShut {NoStop}%
\bibitem [{\citenamefont {Pustogow}(2022)}]{Pustogow2022}%
  \BibitemOpen
  \bibfield  {author} {\bibinfo {author} {\bibfnamefont {A.}~\bibnamefont
  {Pustogow}},\ }\href {https://doi.org/10.3390/solids3010007} {\bibfield
  {journal} {\bibinfo  {journal} {Solids}\ }\textbf {\bibinfo {volume} {3}},\
  \bibinfo {pages} {93} (\bibinfo {year} {2022})}\BibitemShut {NoStop}%
\bibitem [{\citenamefont {Foury-Leylekian}\ \emph {et~al.}(2018)\citenamefont
  {Foury-Leylekian}, \citenamefont {Ilakovac}, \citenamefont {Bal{\'{e}}dent},
  \citenamefont {Fertey}, \citenamefont {Arakcheeva}, \citenamefont {Milat},
  \citenamefont {Petermann}, \citenamefont {Guillier}, \citenamefont
  {Miyagawa}, \citenamefont {Kanoda}, \citenamefont {Alemany}, \citenamefont
  {Canadell}, \citenamefont {Tomic},\ and\ \citenamefont
  {Pouget}}]{FouryLeylekian2018}%
  \BibitemOpen
  \bibfield  {author} {\bibinfo {author} {\bibfnamefont {P.}~\bibnamefont
  {Foury-Leylekian}}, \bibinfo {author} {\bibfnamefont {V.}~\bibnamefont
  {Ilakovac}}, \bibinfo {author} {\bibfnamefont {V.}~\bibnamefont
  {Bal{\'{e}}dent}}, \bibinfo {author} {\bibfnamefont {P.}~\bibnamefont
  {Fertey}}, \bibinfo {author} {\bibfnamefont {A.}~\bibnamefont {Arakcheeva}},
  \bibinfo {author} {\bibfnamefont {O.}~\bibnamefont {Milat}}, \bibinfo
  {author} {\bibfnamefont {D.}~\bibnamefont {Petermann}}, \bibinfo {author}
  {\bibfnamefont {G.}~\bibnamefont {Guillier}}, \bibinfo {author}
  {\bibfnamefont {K.}~\bibnamefont {Miyagawa}}, \bibinfo {author}
  {\bibfnamefont {K.}~\bibnamefont {Kanoda}}, \bibinfo {author} {\bibfnamefont
  {P.}~\bibnamefont {Alemany}}, \bibinfo {author} {\bibfnamefont
  {E.}~\bibnamefont {Canadell}}, \bibinfo {author} {\bibfnamefont
  {S.}~\bibnamefont {Tomic}},\ and\ \bibinfo {author} {\bibfnamefont {J.-P.}\
  \bibnamefont {Pouget}},\ }\href {https://doi.org/10.3390/cryst8040158}
  {\bibfield  {journal} {\bibinfo  {journal} {Crystals}\ }\textbf {\bibinfo
  {volume} {8}},\ \bibinfo {pages} {158} (\bibinfo {year} {2018})}\BibitemShut
  {NoStop}%
\bibitem [{\citenamefont {Liebman}\ \emph {et~al.}(2024)\citenamefont
  {Liebman}, \citenamefont {Miyagawa}, \citenamefont {Kanoda},\ and\
  \citenamefont {Drichko}}]{Liebman2024}%
  \BibitemOpen
  \bibfield  {author} {\bibinfo {author} {\bibfnamefont {J.}~\bibnamefont
  {Liebman}}, \bibinfo {author} {\bibfnamefont {K.}~\bibnamefont {Miyagawa}},
  \bibinfo {author} {\bibfnamefont {K.}~\bibnamefont {Kanoda}},\ and\ \bibinfo
  {author} {\bibfnamefont {N.}~\bibnamefont {Drichko}},\ }\Eprint
  {https://arxiv.org/abs/2403.02676} {arXiv:2403.02676}  (\bibinfo {year}
  {2024})\BibitemShut {NoStop}%
\bibitem [{\citenamefont {Suzuki}\ \emph {et~al.}(2004)\citenamefont {Suzuki},
  \citenamefont {Yamamoto},\ and\ \citenamefont {Yakushi}}]{Suzuki_2004}%
  \BibitemOpen
  \bibfield  {author} {\bibinfo {author} {\bibfnamefont {K.}~\bibnamefont
  {Suzuki}}, \bibinfo {author} {\bibfnamefont {K.}~\bibnamefont {Yamamoto}},\
  and\ \bibinfo {author} {\bibfnamefont {K.}~\bibnamefont {Yakushi}},\ }\href
  {https://doi.org/10.1103/physrevb.69.085114} {\bibfield  {journal} {\bibinfo
  {journal} {Physical Review B}\ }\textbf {\bibinfo {volume} {69}},\ \bibinfo
  {pages} {085114} (\bibinfo {year} {2004})}\BibitemShut {NoStop}%
\end{thebibliography}%

\end{document}